\documentclass[
  10pt,
  longbibliography,
  twocolumn,
  twoside,
]{article}

\usepackage[numbers,round,comma,sort&compress]{natbib}
\setcitestyle{square}

\usepackage[
  left=0.65in,
  right=0.65in,
  top=0.65in,
  bottom=0.65in,
]{geometry}

\usepackage{sectsty}
\sectionfont{\raggedright\large\bfseries}
\subsectionfont{\raggedright\large}

\usepackage[font=small,labelfont=bf]{caption}

\usepackage{fancyhdr}
\usepackage[realmainfile]{currfile}

\usepackage{environ}

\usepackage[export]{adjustbox}

\usepackage{colortbl}

\PassOptionsToPackage{hyphens}{url}\usepackage{hyperref}
\makeatletter
\g@addto@macro{\UrlBreaks}{\UrlOrds}
\makeatother

\hypersetup{
  colorlinks=true,
  allcolors=todoblue,
  urlcolor=todoblue,
  citecolor=todoblue,
  pdfborder={0 0 0},
  breaklinks=true,
}

\usepackage[normalem]{ulem}

\usepackage{soul}

\makeatletter
\let\ftype@table\ftype@figure
\makeatother

\usepackage{array}

\usepackage{color}

\definecolor{todoblue}{RGB}{0, 91, 187}

\newenvironment{textblock}{\renewcommand{\item}{}}{}

\usepackage{changepage}

\usepackage{graphicx}
\usepackage{verbatim}
\usepackage{amssymb}
\usepackage{amsmath}
\usepackage{ifthen}

\usepackage{longtable}

\usepackage{xcolor}

\definecolor{olivegreen}{rgb}{0.33333,.41961,0.18431}
\definecolor{forestgreen}{rgb}{0.13333,.5451,0.13333}

\definecolor{lightgrey}{rgb}{0.7,0.7,0.7}
\definecolor{verylightgrey}{rgb}{0.90,0.90,0.90}
\definecolor{veryverylightgrey}{rgb}{0.95,0.95,0.95}
\definecolor{grey}{rgb}{0.5,0.5,0.5}
\definecolor{darkgrey}{rgb}{0.3,0.3,0.3}
\definecolor{verydarkgrey}{rgb}{0.15,0.15,0.15}

\definecolor{headerblue}{HTML}{33367E}
\definecolor{unitednationsblue}{HTML}{4D88FF}

\definecolor{charcoal}{HTML}{36454F}
\definecolor{cinerous}{HTML}{98817B}
\definecolor{feldgrau}{HTML}{4D5D53}
\definecolor{glaucous}{HTML}{6082B6}
\definecolor{arsenic}{HTML}{3B444B}
\definecolor{xanadu}{HTML}{738678}

\definecolor{firebrick}{HTML}{B22222}
\definecolor{orangered}{HTML}{FF4500}
\definecolor{tomato}{HTML}{FF6347}

\definecolor{orange}{RGB}{255,116,0}

\definecolor{purpletaupe}{HTML}{3B444B}

\definecolor{rose}{HTML}{E3242B}

\colorlet{editnotecolor}{rose}

\definecolor{headerorange}{RGB}{255,116,0}
\definecolor{headergray}{RGB}{230,230,230}

\definecolor{headerpop}{RGB}{230,230,230}

\definecolor{magmalight}{RGB}{252,251,195}
\definecolor{magmalightalt}{RGB}{250,240,184}
\definecolor{magmamedium}{RGB}{245,200,146}
\definecolor{magmadark}{RGB}{224,106,98}

\definecolor{icelight}{RGB}{223,242,244}
\definecolor{icelightalt}{RGB}{189,222,226}
\definecolor{icemedium}{RGB}{132,184,204}
\definecolor{icedark}{RGB}{103,153,191}

\definecolor{traitrowcolor}{RGB}{223,242,244}
\definecolor{traitrowcoloralt}{RGB}{189,222,226}

\definecolor{characterrowcolor}{RGB}{252,251,195}
\definecolor{characterrowcoloralt}{RGB}{250,240,184}

\definecolor{archetyperowcolor}{RGB}{255,213,212} 
\definecolor{archetyperowcoloralt}{RGB}{255,182,179} 

\definecolor{datasetrowcolor}{RGB}{232,244,234}
\definecolor{datasetrowcoloralt}{RGB}{210,231,214}

\NewEnviron{excerpt}{
    \begin{quote}
      \medskip
      \BODY
      \medskip
    \end{quote}
}

\NewEnviron{editnote}{
    \begin{quote}
      \color{rose}
      $\blacksquare$
      \BODY
      \medskip
    \end{quote}
}

\newcommand{\command}[1]{
  \lstinline[language={[LaTeX]TeX},basicstyle=\ttfamily]{#1}
}

\newcommand{\editbox}[2]{
}

\newcommand{\editboxwithlatex}[2]{
}

\usepackage{fancyvrb}

\usepackage{tcolorbox}
\tcbuselibrary{skins,breakable,listings,breakable}

\usepackage{tikz}
\usetikzlibrary{shapes}

\tikzstyle{mybox} = [draw=lightblue!70, fill=lightblue!7, very thick,
    rectangle, rounded corners, inner sep=10pt, inner ysep=20pt]

\tikzstyle{editortitle} =[draw=archetyperowcoloralt, fill=archetyperowcoloralt, text=black]

\newcommand\Loadedframemethod{default}
\usepackage[framemethod=\Loadedframemethod]{mdframed}

\mdfsetup{skipabove=\topskip,skipbelow=\topskip}

\tikzstyle{loglinetitle} =[draw=icedark, fill=icemedium!50, text=black]

\newenvironment{loglinebox}[1][]{

  \ifstrempty{#1}%
  {\mdfsetup{%
    frametitle={%
       \tikz[baseline=(current bounding box.east),outer sep=0pt]
        \node[loglinetitle, anchor=east,rectangle]
        {\strut~~#1:~~\strut};}}
  }%
  {\mdfsetup{%
     frametitle={%
       \tikz[baseline=(current bounding box.east),outer sep=0pt]
        \node[loglinetitle,anchor=east,rectangle]
        {\strut~~#1:~~\strut};}}%
   }%
   \mdfsetup{innertopmargin=5pt,linecolor=icedark,%
             linewidth=0.5pt,topline=true,
             frametitleaboveskip=\dimexpr-\ht\strutbox\relax,}
   \begin{mdframed}[backgroundcolor=icelight,nobreak=true]\relax%
     \raggedright
}{\end{mdframed}}

\tikzstyle{abstracttitle} =[draw=magmadark!75, fill=magmamedium!75, text=black]

\newenvironment{abstractbox}[1][]{

  \ifstrempty{#1}%
  {\mdfsetup{%
    frametitle={%
       \tikz[baseline=(current bounding box.east),outer sep=0pt]
        \node[abstracttitle, anchor=east,rectangle]
        {\strut~~#1:~~\strut};}}
  }%
  {\mdfsetup{%
     frametitle={%
       \tikz[baseline=(current bounding box.east),outer sep=0pt]
        \node[abstracttitle,anchor=east,rectangle]
        {\strut~~#1:~~\strut};}}%
   }%
   \mdfsetup{innertopmargin=5pt,linecolor=magmadark,%
             linewidth=0.5pt,topline=true,
             frametitleaboveskip=\dimexpr-\ht\strutbox\relax,}
   \begin{mdframed}[backgroundcolor=magmalight,nobreak=true]\relax%
     \raggedright
}{\end{mdframed}}

\tikzstyle{infotitle} =[draw=darkgrey, fill=lightgrey!50, text=black]

\newenvironment{infobox}[1][]{

  \ifstrempty{#1}%
  {\mdfsetup{%
    frametitle={%
       \tikz[baseline=(current bounding box.east),outer sep=0pt]
        \node[infotitle, anchor=east,rectangle]
        {\strut~~#1:~~\strut};}}
  }%
  {\mdfsetup{%
     frametitle={%
       \tikz[baseline=(current bounding box.east),outer sep=0pt]
        \node[infotitle,anchor=east,rectangle]
        {\strut~~#1:~~\strut};}}%
   }%
   \mdfsetup{innertopmargin=5pt,linecolor=grey,%
             linewidth=0.5pt,topline=true,
             frametitleaboveskip=\dimexpr-\ht\strutbox\relax,}
   \begin{mdframed}[backgroundcolor=lightgrey!25,nobreak=true]\relax%
     \raggedright
}{\end{mdframed}}

\usepackage{enumitem}

\tikzstyle{changelogtitle} =[draw=darkgrey, fill=lightgrey!50, text=black]

\newenvironment{changelogbox}[1][]{

  \ifstrempty{#1}%
  {\mdfsetup{%
    frametitle={%
       \tikz[baseline=(current bounding box.east),outer sep=0pt]
        \node[changelogtitle, anchor=east,rectangle]
        {\strut~~#1:~~\strut};}}
  }%
  {\mdfsetup{%
     frametitle={%
       \tikz[baseline=(current bounding box.east),outer sep=0pt]
        \node[changelogtitle,anchor=east,rectangle]
        {\strut~~#1:~~\strut};}}%
   }%
   \mdfsetup{innertopmargin=5pt,linecolor=grey,%
             linewidth=0.5pt,topline=true,
             frametitleaboveskip=\dimexpr-\ht\strutbox\relax,}
   \begin{mdframed}[backgroundcolor=lightgrey!25,nobreak=true]\relax%
     \raggedright
     \begin{enumerate}[
         noitemsep,
         leftmargin=10pt,
       ]
}{
     \end{enumerate}
   \end{mdframed}
}

\newcommand{\revisioncolor}{forestgreen}
  
\newif\ifhighlightrevisions
\highlightrevisionsfalse  

\ifhighlightrevisions
\newtcolorbox{revisionbar}{
  enhanced jigsaw,
  boxrule=0pt,
  colframe=white,
  colback=white,
  left=0pt,
  right=0pt,
  top=2pt,
  bottom=2pt,
  breakable,
  before upper={%
    \setlength{\parskip}{1\baselineskip plus .1\baselineskip minus .1\baselineskip}%
    \setlength{\parindent}{0pt}%
  },
  before skip=6pt,
  after skip=6pt,
  overlay={%
    \draw[\revisioncolor, line width=2pt]
      ([xshift=-5pt]frame.north west) --
      ([xshift=-5pt]frame.south west);
  }
}
\else
\newenvironment{revisionbar}{}{}  
\fi

\newcommand{\revision}[1]{#1}

\newcommand{\revisionnote}[1]{}

\usepackage{listings}

\usepackage{stringstrings}
\usepackage{readarray}

\def\firstchar#1#2|{#1}
\edef\tbs{\detokenize{\X}}
\edef\tbs{\expandafter\firstchar\tbs|}
\edef\tlb{\detokenize{{}}}
\edef\tlb{\expandafter\firstchar\tlb|}
\edef\tus{\detokenize{_}}
\newcounter{index}
\newcommand\detokenizeplus[1]{%
  \def\temparg{\detokenize{#1}}%
  \getargsC{\temparg}%
  \setcounter{index}{0}%
  \def\prevmacro{F}%
  \whiledo{\value{index} < \narg}{%
    \stepcounter{index}%
    \isnextbyte[q]{\tbs}{\csname arg\roman{index}\endcsname}%
    \if T\theresult%
      \if T\prevmacro\unskip\else\fi%
      \def\prevmacro{T}%
    \else%
      \def\prevmacro{F}%
   \fi%
    \isnextbyte[q]{\tlb}{\csname arg\roman{index}\endcsname}%
    \if T\theresult\unskip\else\fi%
    \isnextbyte[q]{\tus}{\csname arg\roman{index}\endcsname}%
    \if T\theresult\unskip\else\fi%
    \csname arg\roman{index}\endcsname~%
  }%
}

\usepackage{mathtools}

\newboolean{twocolswitch}

\newcommand{\sindex}[1]{}
\newcommand{\nindex}[1]{}

\newcommand{\www}[1]{\url{#1}}

\usepackage{lettrine}

\usepackage{etoolbox,refcount}

\newcounter{countitems}
\newcounter{nextenumeratecount}
\newcommand{\setupcountitems}{%
  \stepcounter{nextenumeratecount}%
  \setcounter{countitems}{0}%
  \preto\item{\stepcounter{countitems}}%
}
\makeatletter
\newcommand{\computecountitems}{%
  \edef\@currentlabel{\number\c@countitems}%
  \label{countitems@\number\numexpr\value{nextenumeratecount}-1\relax}%
}
\newcommand{\nextenumeratecount}{%
  \getrefnumber{countitems@\number\c@nextenumeratecount}%
}
\makeatother
\AtBeginEnvironment{enumerate}{\setupcountitems}
\AtEndEnvironment{enumerate}{\computecountitems}

\usepackage{xfrac}

\newcommand{\probsymbol}{p}

\newcommand{\numbersymbol}{N}

\newcommand{\bigprobsymbol}{P}
\newcommand{\bigrank}{R}

\newcommand{\indexaraw}{1}
\newcommand{\indexbraw}{2}

\newcommand{\indexa}{(\indexaraw)}
\newcommand{\indexb}{(\indexbraw)}

\newcommand{\systemsymbol}{\Omega}
\newcommand{\elementsymbol}{\tau}

\newcommand{\systema}{\systemsymbol^{\indexa}}
\newcommand{\systemb}{\systemsymbol^{\indexb}}

\newcommand{\Ntypesa}{\numbersymbol_{\indexaraw}}
\newcommand{\Ntypesb}{\numbersymbol_{\indexbraw}}

\newcommand{\strongdistance}{\mathcal{D}}

\newcommand{\bigrankordering}{\bigrank_{\indexaraw,\indexbraw;\alpha}}
\newcommand{\bigrankorderingalpha}[1]{\bigrank_{\indexaraw,\indexbraw;#1}}

\newcommand{\probdiv}[1]{D^{\textnormal{P}}_{#1}}
\newcommand{\probdivelement}[1]{\delta D^{\textnormal{P}}_{#1,\elementsymbol}}

\newcommand{\probdivalpha}{\probdiv{\alpha}}

\newcommand{\ptdnorm}{\mathcal{N}_{\indexaraw,\indexbraw;\alpha}^{\textnormal{P}}}
\newcommand{\invptdnorm}{\frac{1}{\ptdnorm}}

\newcommand{\ptdnormalpha}[1]{\mathcal{N}_{\indexaraw,\indexbraw;#1}^{\textnormal{P}}}

\newcommand{\logten}{\textnormal{log}_{10}}

\setboolean{twocolswitch}{true}

\usepackage{natbib}
\setcitestyle{square}
\raggedright

\newcommand{\zenodobaseurl}{\href{https://zenodo.org}{zenodo}}
\newcommand{\zenodorecord}{15006457}
\newcommand{\zenodourl}{https://doi.org/10.5281/zenodo.\zenodorecord}
\newcommand{\zenodolink}{\href{\zenodourl}{\zenodourl/}}
\newcommand{\zenodofileslink}{https://zenodo.org/records/\zenodorecord/files}

\newcommand{\gitlaburl}{https://gitlab.com/compstorylab/allotaxonometer/}
\newcommand{\gitlablink}{\href{\gitlaburl}{\gitlaburl}}

\newcommand{\onlineappendicesurl}{https://compstorylab.org/allotaxonometry/}
\newcommand{\onlineappendiceslink}{\href{\onlineappendicesurl}{\onlineappendicesurl}}
\newcommand{\onlineappendices}{Online Appendices (\onlineappendiceslink)}
\newcommand{\onlineappendicesplain}{Online Appendices}

\newcommand{\arxivonly}[1]{}

\usepackage{authblk}

\begin{document}

\title{
  \textbf{\protect Probability-turbulence divergence:\\
A tunable allotaxonometric instrument \\
for comparing heavy-tailed
\revision{type-probability}
distributions




}

\bigskip

\large
\protect\textit{PLOS Complex Systems},
\textbf{3}(7): e0000077,
2026
\\
\href{https://doi.org/10.1371/journal.pcsy.0000077}{https://doi.org/10.1371/journal.pcsy.0000077}

}


\renewcommand*{\Authsep}{, }
\renewcommand*{\Authand}{, }
\renewcommand*{\Authands}{, }
\renewcommand*{\Affilfont}{\normalsize\normalfont}
\renewcommand*{\Authfont}{}
\setlength{\affilsep}{2em}

\author[1,2,3,4]{Peter~Sheridan~Dodds\thanks{peter.dodds@uvm.edu}}
\author[5]{Joshua~R.~Minot}
\author[1,2]{Michael~V.~Arnold}
\author[6,7]{Thayer~Alshaabi}
\author[8]{Jane~Lydia~Adams}
\author[5]{Andrew~J.~Reagan}
\author[1,2,9]{Christopher~M.~Danforth}


\affil[1]{
  Computational Story Lab,
  Vermont Advanced Computing Center,
  University of Vermont,
  Burlington,
  VT 05405,
  USA
}

\affil[2]{
  Vermont Complex Systems Institute,
  MassMutual Center of Excellence for Complex Systems and Data Science,
  University of Vermont,
  Burlington,
  VT 05405,
  USA
  }

\affil[3]{
  Department of Computer Science,
  University of Vermont,
  Burlington,
  VT 05405,
  USA
}

\affil[4]{
  Santa Fe Institute,
  1399 Hyde Park Rd,
  Santa Fe,
  NM 87501,
  USA
}

\affil[5]{
  MassMutual Data Science,
  Amherst,
  MA 01002,
  USA
}

\affil[6]{
  Howard Hughes Medical Institute,
  Janelia Research Campus,
  Ashburn,
  VA 20147,
  USA
} 

\affil[7]{
  Advanced Bioimaging Center,
  University of California Berkeley,
  Berkeley,
  CA 94720,
  USA
}

\affil[8]{
  Data Visualization Lab,
  Khoury College of Computer Sciences,
  Northeastern University,
  Boston,
  MA 02115,
  USA
}

\affil[9]{
  Department of Mathematics \& Statistics,
  University of Vermont,
  Burlington,
  VT 05405,
  USA
  }

\date{\today}

\maketitle


\mbox{}

\bigskip
\bigskip
\bigskip

\hspace*{-230pt}
\begin{minipage}{420pt}

  \begin{tabular}{>{\raggedright\arraybackslash}p{240pt}p{30pt}p{180pt}}

    \parbox{240pt}{    
      \begin{loglinebox}[Logline]
        \raggedright
        Probability-turbulence divergence is an allotaxonometric instrument
designed to compare complex systems comprised of many element types
which follow heavy-tailed abundance distributions.

\medskip

Allotaxonographs for probability-turbulence divergence provide map-and-list
visualizations that:
(1)~Properly accommodate zero probabilities,
and
(2)~Surface which elements
most differentiate the composition of any two systems.

\medskip

Constructed with a single parameter $\alpha$,
probability-turbulence divergence can, in the manner of a physical instrument,
be `tuned', foregrounding types from
rare ($\alpha=0$)
to
common ($\alpha=\infty$),
and offers the possibility of a scale-equalizing $\alpha$.

\medskip


Probability-turbulence divergence generalizes and unifies a wide
range of existing distance measures.

        \smallskip
      \end{loglinebox}
    }

    &
    
    &
    
    \parbox{180pt}{    
      \includegraphics[width=180pt,valign=m,frame]{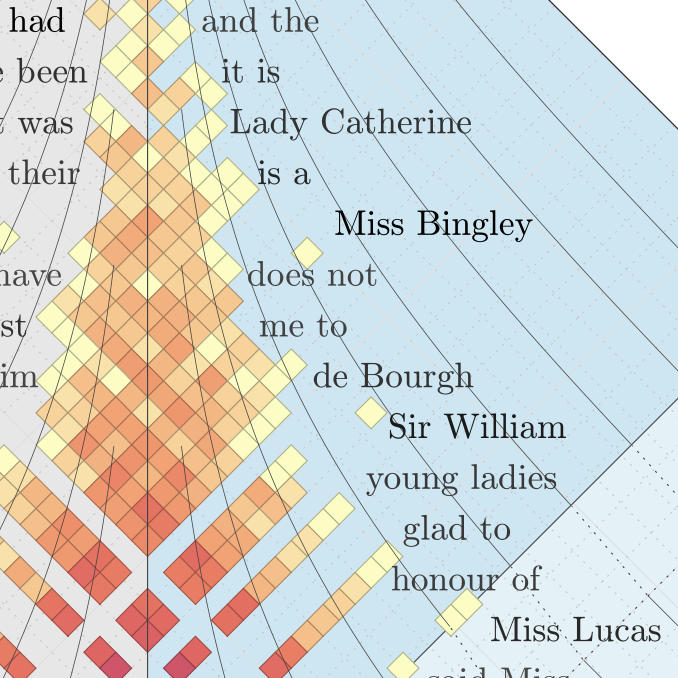}
    }
    
  \end{tabular}
  
\end{minipage}

\clearpage

\newgeometry{
  left=2in,
  right=2in,
  top=1in,
  bottom=1in,
  }

\onecolumn

\renewcommand{\baselinestretch}{1.25}
\selectfont


\begin{abstractbox}[Abstract]
  \raggedright
  \begin{textblock}
\item
  Real-world complex systems often comprise many distinct types of elements as well
  as many more types of networked interactions between elements.
\item
  When the relative abundances of types can be measured well,
  we often observe heavy-tailed distributions for
  type probabilities (or relative rates).
\item
  For the comparison of \revision{type-probability distributions} of two systems
  or a system with itself at different
  points in time---a facet of allotaxonometry---a great range of
  probability divergences are available.
  \\
  \medskip
\item 
  Here, we introduce and explore `probability-turbulence divergence',
  a tunable, straightforward,
  and interpretable instrument for comparing normalizable \revision{type-probability distributions.}
\item 
  We model probability-turbulence divergence (PTD) after rank-turbulence divergence (RTD).
\item 
  While probability-turbulence divergence
  is more limited in application
  than rank-turbulence divergence,
  it is more sensitive
  to changes in type probability.
  \\
  \medskip
\item
  We show how probability-turbulence divergence
  either explicitly or functionally generalizes
  many existing kinds of distances and measures,
  including, as special cases,
  $L_{q}$ norms,
  the S{\o}rensen-Dice coefficient
  (the $F_{1}$ statistic),
  and 
  the Hellinger distance.
\item
  We discuss similarities
  with the generalized entropies of R{\'e}nyi and Tsallis,
  and the diversity indices (or Hill numbers) from ecology.
  \\
  \medskip
\item
  We then build allotaxonographs to display probability turbulence,
  incorporating a way to visually
  accommodate zero probabilities for `exclusive types'
  which are types that appear in only one system.
\item
  \revision{Using flipbooks, we show how tuning PTD's single parameter
  informs the user how two systems diverge for types that
  are rare, common, and at all scales in between.}
\item 
  \revision{We demonstrate that PTD can be tuned to a `scale-equalizing'
  view that is non-universal and dependent on the systems
  being compared.}
  \\
  \medskip
\item
  We explore comparisons of example distributions
  taken from
  literature,
  social media,
  and ecology.
  \\
  \medskip
\item
  We close with thoughts on open problems
  concerning  the 
  optimization of the tuning of rank- and
  probability-turbulence divergence.
\end{textblock}


  \smallskip
\end{abstractbox}


\begin{infobox}[Keywords]
  \centering
  complex~systems;
dynamical~systems;
size-rank~distributions;
Zipf~distributions;
heavy-tailed~distributions;
allotaxonometry;
allotaxonometer;
allotaxonograph;
divergence;
lexical~calculus;
lexical~turbulence;
type~calculus;
type~turbulence;
probability-turbulence~divergence;
measurement;
Jane~Austen;
language;
social~media;
literature;
ecology

 \smallskip
\end{infobox}

\renewcommand{\baselinestretch}{1}
\selectfont

\twocolumn

\restoregeometry

\clearpage

\tableofcontents

\clearpage

\renewcommand\floatpagefraction{0}

\revisionnote{
  We add notes about changes to the paper (like this one)
  in red.
  These notes are not part of the paper.
}

\revisionnote{
  We highlight changes to the paper in green,
  either inline or with a bar in the margin for a larger section.
  We do not highlight minor copyedits.
}


\revisionnote{
  We have adjusted ``heavy-tailed categorical distributions''
  to ``heavy-tailed type-probability distributions''
  in the title and throughout.
}

\revisionnote{
  Wherever ambiguous, we have now made clear that frequency is normalized frequency,
  and that the sum of normalized frequencies over all times is 1.
  For convenience, we only use probability in our equations, and we have noted this.
}

\section{Introduction}
\label{sec:probturbdiv.intro}

\begin{textblock}
\item
  Driven by an interest in developing allotaxonometry~\cite{dodds2023a}---the
  detailed comparison of any two complex systems comprising many types of elements---we
  often find we need to compare `normalized-size-rank' distributions:
\item 
  Heavy-tailed categorical distributions of normalized type-sizes~\cite{zipf1949a,simon1955a,newman2005b,clauset2009b}, 
\item
  often referred to as Zipf distributions.
\item
  The normalized sizes we are most interested in
  are probabilities or relative rates.
\item
  We take a relaxed definition of what a heavy tail means for a size-rank distribution:
  A slow decay over orders of magnitude in type rank.
\item
  Though not required, power-law decay tails are emblematic signatures
  of heavy-tailed distributions commonly presented by 
\item
  complex systems~\cite{barabasi1999a,mitchell2009a,bettencourt2007a,magurran2013a,scholz2019a},
\item 
  both observed and theoretical,
\item 
  and provide important examples to consider in our efforts to build a comparison tool.
\end{textblock}

\begin{textblock}
\item
  Across fields,
  efforts to measure and explain how two probability distributions differ
  have led to the generation of a great many
  probability divergences~\cite{deza2006a,cha2007a,cichocki2010a,monroe2008a}.
\item 
  Divergences have been developed
  for a host of motivations quite apart from our focus here on allotaxonometry,
  with example families scaffolded around
  $L^{p}$-norms,
  inner products,
  and information-theoretic constructions.
\item 
  As we will discuss, for heavy-tailed distribution comparisons which exhibit variable
  `probability turbulence'~\cite{pechenick2017a,dodds2023a},
  we find these divergences lack appropriate adaptability.
\end{textblock}

\begin{textblock}
\item
  Here, we introduce a tunable, interpretable instrument that we call
  probability-turbulence divergence (PTD) along with related allotaxonographs---visualizations
  which show in detail how two categorical distributions differ according to a given measure.
\item 
  We refer the reader to Ref.~\cite{dodds2023a} for our motivation
  for creating allotaxonometry and allotaxonographs, the notion of rank turbulence,
  and a detailed justification for the form of rank-turbulence divergence (RTD)
  we developed there.
\item 
  We establish probability-turbulence divergence using largely the same arguments,
  \revision{though PTD is distinct from RTD in its application and properties.}
\item 
  We will therefore be concise in our presentation and expand as needed when
  the probability version's behavior departs from that of its rank counterpart.
\end{textblock}

\begin{textblock}
\item
  In Sec.~\ref{sec:probturbdiv.probdiv},
  we formally define probability-turbulence divergence.
\item
  We describe the divergence's general analytic behavior
  as a function of its single parameter, $\alpha$,
  and we determine its form
  for the two limits of the parameter, $\alpha$=0 and $\alpha$=$\infty$.
\item
  When $\alpha$=$0$, in particular, we find an interesting departure
  from the equivalent tuning for rank-turbulence divergence,
\item
  and which we will later connect to
  the S{\o}rensen-Dice coefficient~\cite{dice1945a,sorensen1948a}
  and the $F_{1}$ score~\cite{vanrijsbergen1979a}.
\end{textblock}

\begin{textblock}
\item
  In Sec.~\ref{sec:probturbdiv.links}, we show that probability-turbulence
  divergence is either a generalization of or may be connected to a 
  number of other kinds of divergences and similarities (e.g., the S{\o}rensen-Dice coefficient),
\item
  and then discuss limited functional similarities with
  the
  R{\'e}nyi entropy and diversity indices~\cite{renyi1961a,tsallis2001a,keylock2005a,hill1973a,jost2006a}.
\end{textblock}

\begin{textblock}
\item
  In Sec.~\ref{sec:probturbdiv.graphicalinstrument},
  we then provide realizations of probability-turbulence divergence as an instrument
  through example allotaxonographs.
\item 
  For three disparate examples, we consider
\item 
  1. Normalized frequency of $n$-gram use in Jane Austen's Pride and Prejudice,
\item 
  2. Normalized frequency of $n$-gram use on Twitter,
\item 
  and
  3. Tree species abundance~\cite{condit2019a}.
\item 
  We show how, for the kinds of heavy-tailed distributions we are interested in,
  a probability-turbulence divergence histogram can be constructed to
  accommodate both a logarithmic scale and the presence of zero probabilities.
\item 
  Similar to rank-turbulence divergence histograms, these graphs clearly
  show whether or not a probability-based divergence is a suitable choice
  for any given comparison.
\item 
  We explore allotaxonographs for the full range of parameter $\alpha$,
  highlighting the existence of
  special, non-universal `scale-equalizing' values of $\alpha$.
\item 
  \revision{To demonstrate the tunability of probability-turbulence divergence,
  we provide allotaxonographs `flipbooks' as
  part of the supplementary information~\cite{dodds_PTD_flipbooks_2025}.}
\item 
  \arxivonly{See also \onlineappendicesplain.}
\item 
  \revision{We consider the construction and examination of these
    flipbooks essential to our
    method of comparison for any two systems.}
\end{textblock}

\begin{textblock}
\item
  We outline data, allotaxonometry code, and supplementary material
  in Sec.~\ref{sec:probturbdiv.methods},
\item
  and we offer some concluding thoughts
  in Sec.~\ref{sec:probturbdiv.concludingremarks}.
\end{textblock}

\begin{table}[t]
  \centering
  \begin{minipage}{0.9\linewidth}

\begin{loglinebox}[Glossary of Key Terms]
  \begin{textblock}
  \item
    \textbf{Allotaxonometry:}
    The comparison of any two complex systems comprising many different types of elements,
    likely of highly varying `size' (e.g., number, probability)
    and compositional structure (e.g., hierarchies, networks).
  \item 
    Coined from Greek roots:
    allo- (`other'),
    taxis (`order/arrangement'),
    and metron (`measure').
    Introduced in Ref.~\cite{dodds2023a}.
  \item
    \smallskip
    \newline
    \textbf{Allotaxonograph:} A map-and-list visualization
    of how types (e.g., words, species)
    differ in size (e.g., rank, abundance) between two systems.
  \item 
    \smallskip
    \newline
    \textbf{Probability-Turbulence Divergence (PTD):} A tunable divergence measure introduced in this paper for comparing two probability distributions over categorical types. PTD generalizes similarity measures (e.g., S{\o}rensen-Dice) by incorporating a parameter $\alpha$ that adjusts the weight given to rare vs. common types.
  \item
    \smallskip
    \newline
    \textbf{Rank-Turbulence Divergence (RTD):}
    An instrument for comparing two ranked lists that come from heavy-tailed distributions of type sizes,
    which need not be known (i.e., only ranks matter).
  \item
    \smallskip
    \newline
    \textbf{Tuning parameter $\alpha$:}
    Real-valued parameter that is $\ge 0$ which controls how PTD (or RTD) emphasizes rare vs.\ common types.
    Lower values highlight rare types; higher values prioritize common ones.
    The parameter can also be viewed as a diagnostic quantity.
  \item
    \smallskip
    \newline
    \textbf{Scale-equalizing $\alpha$:}
    Where probability turbulence scales well, a value of $\alpha$ for which rare and common types and types in between
    all contribute
    in a balanced way to probability-turbulence divergence.
    The value of $\alpha$ is not exact but rather represents
    a range outside of which rare or common types
    begin to dominate the divergence score.
  \item
    \smallskip
    \newline
    \textbf{Flipbook sweeps over $\alpha$:}
    PDF booklets of allotaxonographs for $\alpha$ systematically varying from 0 to $\infty$ (see Eq.~\ref{eq:probturbdiv.alphavals}).
    Allows users to observe how the map-and-list visualizations change 
    with $\alpha$, and to determine, if desired and if possible, a scale-equalizing value of $\alpha$.
  \end{textblock}
\end{loglinebox}

  \end{minipage}
  \caption{
    Glossary.
  }
  \label{tab:probturbdiv.glossary}
\end{table}

\section{Probability-turbulence divergence}
\label{sec:probturbdiv.probdiv}


\begin{textblock}
\item
  We aim to compare two systems $\systemsymbol_{\indexaraw}$
  and
  $\systemsymbol_{\indexbraw}$
  for both of which we have a list of component types
  and their probabilities.
\item
  For simplicity, we will use probability
  in our general derivations.
\item
  \revision{
    In describing any particular distribution,
    we will use an appropriate terminology,
    such as 
    normalized frequency, relative frequency, or rate of usage.
    }
\item
  All normalizations must be such that the sum of `sizes' is 1.
\end{textblock}

\begin{textblock}
\item
  We denote a type by $\elementsymbol$ and its probability
  in the two systems as 
  $\probsymbol_{\elementsymbol,\indexaraw}$
  and
  $\probsymbol_{\elementsymbol,\indexbraw}$.
\item
  We represent the probability distributions for the two systems
  as
  $\bigprobsymbol_{\indexaraw}$
  and
  $\bigprobsymbol_{\indexbraw}$.
\item
  We call types that are present in one system only `exclusive types'.
\item
  We will use expressions of the form
  $\systema$-exclusive and $\systemb$-exclusive to indicate 
  to which system a type solely belongs.
\end{textblock}

\begin{textblock}
\item
  We are interested in divergences that are
  some function of a sum of contributions by type.
\item
  Here, we will consider a single-parameter family of divergences that are
  of the simplest form, i.e., a direct sum
  of contributions:
\item 
\begin{gather}
  \probdivalpha(
  \bigprobsymbol_{\indexaraw}
  \,\Vert\,
  \bigprobsymbol_{\indexbraw}
  )
  =
  \sum_{\elementsymbol \in \bigrankordering}
  \probdivelement{\alpha}(
  \bigprobsymbol_{\indexaraw}
  \,\Vert\,
  \bigprobsymbol_{\indexbraw}
  ).
  \label{eq:probturbdiv.probturbdiv_sum}
\end{gather}
\item
  By the rank-ordered set $\bigrankordering$,
  we indicate the union of all types from both systems,
  sequenced such that the contributions
\end{textblock}
$
\probdivelement{\alpha}(
\bigprobsymbol_{\indexaraw}
\,\Vert\,
\bigprobsymbol_{\indexbraw}
)
$
\begin{textblock}
\item
  are monotonically decreasing (hence the necessity of an $\alpha$ subscript).
\item
  We impose this order for general good housekeeping, secondarily
  allowing  us to handle possibilities such as truncated summations
  due to sampling, or convergence issues for theoretical examples.
\end{textblock}
 
\begin{revisionbar}
  
\begin{textblock}
\item
  Our motivation starts with the observed phenomenon of type turbulence
  in the comparison of complex systems with heavy-tailed distributions
  of types according to some kind of size~\cite{pechenick2017a,dodds2023a}.
\item 
  Our objective is to create a tunable probability divergence measure
  that can contend with variable type turbulence.
\item
  In doing so, we will incidentally arrive at a divergence that
  can be tuned from depending on only the most rare words
  to only the most common words.
\end{textblock}

\begin{textblock}
\item
  As a start, and in a manner similar to how we approached
  rank-turbulence divergence~\cite{dodds2023a},
  we consider the base difference quantity for type $\elementsymbol$:
\item
  \begin{align}
    \Bigm\lvert    
    \left[\,\probsymbol_{\elementsymbol,\indexaraw}\right]^{\alpha}
    -
    \left[\,\probsymbol_{\elementsymbol,\indexbraw}\right]^{\alpha}
    \Bigm\lvert,
    \label{eq:probturbdiv.probdiv_core}
  \end{align}
\item
  where $\alpha \ge 0$ is a tuning parameter.
\item
  Evidently, low $\alpha$ will dampen differences between probabilities
  and high $\alpha$ will accentuate them.
\item
  However,
  increasing $\alpha$ will also reduce the
  difference quantity to 0,
  removing its ability to contribute.
\item
  We partly solve this by raising the quantity to $1/\alpha$:
\item
  \begin{align}
  \Bigm\lvert    
  \left[\,\probsymbol_{\elementsymbol,\indexaraw}\right]^{\alpha}
  -
  \left[\,\probsymbol_{\elementsymbol,\indexbraw}\right]^{\alpha}
  \Bigm\lvert^{1/(\alpha)}.    
  \label{eq:probturbdiv.probdiv_core_mod}
\end{align}
\item
  Now the large $\alpha$ limit functions well with
  $
  \lim_{\alpha \rightarrow \infty}
  \lvert    
  \left[\,\probsymbol_{\elementsymbol,\indexaraw}\right]^{\alpha}
  -
  \left[\,\probsymbol_{\elementsymbol,\indexbraw}\right]^{\alpha}
  \lvert
  =
  \max_{\elementsymbol}
  (
  \probsymbol_{\elementsymbol,\indexaraw},
  \probsymbol_{\elementsymbol,\indexbraw}
  ).
  $
\item
  But now we have a problem for the $\alpha \rightarrow 0$ limit
  as the difference quantity will tend toward $\infty$
  unless in the rare case that the probabilities are the same.
\item
  We adjust the difference quantity once more in a parsimonious way:
\item
  \begin{align}
    \frac{\alpha+1}{\alpha}
    \sum_{\elementsymbol \in \bigrankordering}
    \Bigm\lvert
    \left[\,\probsymbol_{\elementsymbol,\indexaraw}\right]^{\alpha}
    -
    \left[\,\probsymbol_{\elementsymbol,\indexbraw}\right]^{\alpha}
    \Bigm\lvert^{1/(\alpha+1)}.
    \label{eq:probturbdiv.probdiv_core_final}
  \end{align}
\item
  The behavior for large $\alpha$ remains the same,
  and now for $\alpha \rightarrow 0$,
  the difference quantity converges and does so in a meaningful way.
\item
  We explicate the $\alpha \rightarrow 0$ limit
  fully below in Sec.~\ref{subsec:probturbdiv.probturbdivzero}.
\end{textblock}

\end{revisionbar}

\begin{textblock}
\item
  We now define probability-turbulence divergence as:
\item 
\begin{align}
  &
  \probdivalpha(
  \bigprobsymbol_{\indexaraw}
  \,\Vert\,
  \bigprobsymbol_{\indexbraw}
  )
  \nonumber
  \\
  &
  =
  \invptdnorm
  \frac{\alpha+1}{\alpha}
  \sum_{\elementsymbol \in \bigrankordering}
  \Bigm\lvert
  \left[\,\probsymbol_{\elementsymbol,\indexaraw}\right]^{\alpha}
  -
  \left[\,\probsymbol_{\elementsymbol,\indexbraw}\right]^{\alpha}
  \Bigm\lvert^{1/(\alpha+1)}.
  \label{eq:probturbdiv.probdiv_definition}
\end{align}
\item
  where
  the parameter $\alpha$ may be tuned from 0 to $\infty$
  and
  $\ptdnorm$
  is a normalization factor.
\item
  Per Ref.~\cite{dodds2023a} and below
  in Sec.~\ref{subsec:probturbdiv.probturbdivzero},
\item
  the roles of the prefactor $(\alpha+1)/\alpha$
  and the power $1/(\alpha+1)$ are to govern the behavior
  of PTD in the limit $\alpha \rightarrow 0$.
\end{textblock}

\begin{revisionbar}
    \begin{textblock}
  \item
    As we will show below,
    sweeping across $\alpha$
    allows the user to accentuate the importance
    of the most rare types (when $\alpha = 0$)
    through to the most common types ($\alpha = \infty$).
  \item
    Further, for certain special values of $\alpha$,
    PTD corresponds in behavior to many extant divergences and differences.
  \item
    And finally, for system comparisons that exhibit scaling of probability-turbulence,
    users will be able to find a scale-equalizing $\alpha$.
  \end{textblock}
\end{revisionbar}

\begin{textblock}
\item
  By construction and regardless of the choice of
  normalization factor,
  we can see from
  Eq.~\ref{eq:probturbdiv.probdiv_definition}
  that probability-turbulence divergence will equal 0
  when both distributions are the same.
\item
  (We show below that for  $\alpha = 0$, distinct distributions can
  also register
\item 
  $
  \probdivalpha(
  \bigprobsymbol_{\indexaraw}
  \,\Vert\,
  \bigprobsymbol_{\indexbraw}
  )
  =0$.)
\end{textblock}

\begin{textblock}
\item
  The core of~Eq.~\ref{eq:probturbdiv.probdiv_definition}
  is the absolute value of the difference of each type $\elementsymbol$'s
  probability raised to the power of $\alpha$:
\item 
\begin{align}
  \Bigm\lvert
  \left[\,\probsymbol_{\elementsymbol,\indexaraw}\right]^{\alpha}
  -
  \left[\,\probsymbol_{\elementsymbol,\indexbraw}\right]^{\alpha}
  \Bigm\lvert.
  \label{eq:probturbdiv.probdiv_definition_core}
\end{align}
\item
  This $\alpha$-tuned quantity controls the
  order of contributions by types to the overall value of PTD.
\item
  As $\alpha \rightarrow 0$,
  lower probabilities---corresponding to the rare types---are relatively accentuated.
  For $\alpha \rightarrow \infty$,
\item
  the higher of the two probabilities will dominate (unless they are equal),
\item
  meaning the most common types will come to the fore.
\end{textblock}

\subsection{Normalization for probability-turbulence divergence}
\label{subsec:probturbdiv.normalization}

\begin{textblock}
\item
  As for rank-turbulence divergence, we choose $\ptdnorm$
  so that when the two systems are entirely disjoint---that is, they share
  no types---then probability-turbulence divergence maximizes at 1.
\item
  The normalization is thus specific to the two distributions being
  compared.
\item 
  We imagine that the types in each system have an extra descriptor
  specifying belonging to
\item 
  $\systema$
  or
  $\systemb$.
\item 
  With no matching types, the probability of a type present in one system is zero
  in the other, 
\item
  and  the sum can be split between the two systems' types:
\item 
\begin{align}
  &
  \ptdnorm
  =
  \nonumber
  \\
  &
  \frac{\alpha+1}{\alpha}
  \sum_{\elementsymbol \in \bigrank_{\indexaraw}}
  \left[\,
  \probsymbol_{\elementsymbol,\indexaraw}
  \right]^{\alpha/(\alpha+1)}
  +
  \frac{\alpha+1}{\alpha}
  \sum_{\elementsymbol \in \bigrank_{\indexbraw}}
  \left[\,
  \probsymbol_{\elementsymbol,\indexbraw}
  \right]^{\alpha/(\alpha+1)},
  \label{eq:probturbdiv.probdiv_norm_basic}
\end{align}
\item
  where
  $\bigrank_{\indexaraw}$
  and
  $\bigrank_{\indexbraw}$
\item
  are the rank-ordered sets of types for each system.
\end{textblock}

\begin{textblock}
\item
  We can more compactly express the normalization as:
\item
  \begin{align}
    \ptdnorm
    =
    \frac{\alpha+1}{\alpha}
    \sum_{\elementsymbol \in \bigrankordering}
    \left(
    \left[\,
      \probsymbol_{\elementsymbol,\indexaraw}
      \right]^{\alpha/(\alpha+1)}
    +
    \left[\,
      \probsymbol_{\elementsymbol,\indexbraw}
      \right]^{\alpha/(\alpha+1)}
    \right),
    \label{eq:probturbdiv.probdiv_norm}
  \end{align}
\item
  where,
  as for the definition of probability-turbulence divergence in Eq.~\ref{eq:probturbdiv.probdiv_definition},
\item 
  the sum for the normalization is again over the ordered set
  $\bigrankordering$.
\item 
  Types that appear in both systems will have their contribution
\item 
  $
  \left[\,
    \probsymbol_{\elementsymbol,\indexaraw}
    \right]^{\alpha/(\alpha+1)}
  $
\item 
  and
\item 
  $
  \left[\,
    \probsymbol_{\elementsymbol,\indexbraw}
    \right]^{\alpha/(\alpha+1)}
  $
  counted appropriately.
\end{textblock}

\subsection{\revision{Normalization for $\alpha$=$0$}}
\label{subsec:probturbdiv.probturbdivzero}

\begin{textblock}
\item
  The $\alpha \rightarrow 0$ limit requires some care and will
  vary from the equivalent limit for rank-turbulence divergence~\cite{dodds2023a}.
\item
  First, at the level of individual type contribution, if
  both
  $\probsymbol_{\elementsymbol,\indexaraw} > 0$
  and
  $\probsymbol_{\elementsymbol,\indexbraw} > 0$
  then
\item
  \begin{equation}
    \lim_{\alpha \rightarrow 0}
    \frac{\alpha+1}{\alpha}
    \Bigm\lvert
    \left[\,\probsymbol_{\elementsymbol,\indexaraw}\right]^{\alpha}
    -
    \left[\,\probsymbol_{\elementsymbol,\indexbraw}\right]^{\alpha}
    \Bigm\lvert^{1/(\alpha+1)}
    =
    \left\lvert
    \textnormal{ln}
    \,
    \frac{\probsymbol_{\elementsymbol,\indexbraw}
    }{\probsymbol_{\elementsymbol,\indexaraw}}
    \right\rvert.
    \label{eq:probturbdiv.elementprobturbdiv_zerolimit1}
  \end{equation}
\item
  If instead a type $\elementsymbol$ is exclusive to
  one system, meaning either
  $\probsymbol_{\elementsymbol,\indexaraw} = 0$
  or
  $\probsymbol_{\elementsymbol,\indexbraw} = 0$,
\item
  then the limit diverges as $1/\alpha$, which would seem problematic.
\item
  We nevertheless will arrive at a well-behaved divergence
  through the normalization term $\ptdnormalpha{0}$.
\end{textblock}

\begin{textblock}
\item
  Requiring as we have that the extreme of disjoint systems have a divergence of 1,
\item 
  we observe that each of the types in the case of
  disjoint systems would contribute $1/\alpha$.
\item 
  Therefore, in the $\alpha \rightarrow 0$ limit, we must have:
\item
  \begin{equation}
    \ptdnormalpha{\alpha}
    \rightarrow
    \frac{1}{\alpha}
    \left(
    \Ntypesa + \Ntypesb
    \right).
    \label{eq:probturbdiv.elementprobturbdiv_zerolimit_norm}
  \end{equation}
\item
  Because the normalization also diverges as $1/\alpha$, the divergence
  will be zero when there are no exclusive types and non-zero when
  there are exclusive types.
\item
  We can combine these cases into a single expression:
\item 
  \begin{align}
    \probdiv{0}(
    \bigprobsymbol_{\indexaraw}
    \,\Vert\,
    \bigprobsymbol_{\indexbraw}
    )
    =
    \frac{1}{
      \left(
      \Ntypesa + \Ntypesb
      \right)
    }
    \sum_{\elementsymbol \in \bigrankorderingalpha{0}}
    \left(
    \delta_{\probsymbol_{\elementsymbol,\indexaraw},0}
    +
    \delta_{0,\probsymbol_{\elementsymbol,\indexbraw}}
    \right).
    \label{eq:probturbdiv.probdiv_alpha_zero_exclusive}
  \end{align}
\item
  The term
  $
  \left(
  \delta_{\probsymbol_{\elementsymbol,\indexaraw},0}
  +
  \delta_{0,\probsymbol_{\elementsymbol,\indexbraw}}
  \right)
  $
\item
  returns 1 if either
\item
  $\probsymbol_{\elementsymbol,\indexaraw} = 0$
  or
  $\probsymbol_{\elementsymbol,\indexbraw} = 0$,
\item
  and
  0 otherwise when both
\item
  $\probsymbol_{\elementsymbol,\indexaraw} > 0$
  and
  $\probsymbol_{\elementsymbol,\indexbraw} > 0$.
\item
  (By construction, we cannot have
  $
  \probsymbol_{\elementsymbol,\indexaraw}
  =
  \probsymbol_{\elementsymbol,\indexbraw}
  =
  0$
  as each type must be present in either one or both systems.)
\end{textblock}

\begin{textblock}
\item
  We see then that
\item
  $
  \probdiv{0}(
  \bigprobsymbol_{\indexaraw}
  \,\Vert\,
  \bigprobsymbol_{\indexbraw}
  )
  $
\item
  is the ratio of types that are exclusive to one system
  relative to the total possible such types,
\item
  $
  \Ntypesa + \Ntypesb.
  $
\item
  If and only if
  all types appear in both systems with whatever variation in probabilities, then
\item
  $
  \probdiv{0}(
  \bigprobsymbol_{\indexaraw}
  \,\Vert\,
  \bigprobsymbol_{\indexbraw}
  )
  =
  0.
  $
\item
  The limit of $\alpha = 0$ therefore exhibits special behavior.
  For $\alpha > 0$,
\item
  probability-turbulence divergence only scores
  0 for exactly matching distributions.
\end{textblock}

\subsection{Type contribution ordering for the limit of $\alpha$=$0$} 
\label{subsec:probturbdiv.probturbdivzero_contributions}

\begin{textblock}
\item
  In terms of contribution to the divergence score,
  all exclusive types supply
  a weight of
  $1/(\Ntypesa + \Ntypesb)$.
\item
  We can order them by preserving their ordering as $\alpha \rightarrow 0$,
\item
  which amounts to ordering by descending probability in the system
  in which they appear.
\end{textblock}

\begin{textblock}
\item
  And while types that appear in both systems make no contribution
  to
\item
  $
  \probdiv{0}(
  \bigprobsymbol_{\indexaraw}
  \,\Vert\,
  \bigprobsymbol_{\indexbraw}
  ),
  $
\item
  we can still order them according to
  the log ratio of their probabilities,
  Eq.~\ref{eq:probturbdiv.elementprobturbdiv_zerolimit1}.
\end{textblock}

\begin{textblock}
\item
  The overall ordering of types by divergence contribution
  for $\alpha$=$0$ is then:
\item 
  (1) exclusive types by descending
  probability and then
\item 
  (2) types appearing in both systems by
  descending log ratio.
\end{textblock}

\subsection{\revision{Normalization for $\alpha$=$\infty$}}
\label{subsec:probturbdiv.probturbdivinfty}

\begin{textblock}
\item
  The $\alpha \rightarrow \infty$ limit is straightforward and
  in line with that of rank-turbulence divergence~\cite{dodds2023a}:
\item
  \begin{equation}
    \probdiv{\infty}(
    \bigprobsymbol_{\indexaraw}
    \,\Vert\,
    \bigprobsymbol_{\indexbraw}
    )
    =
    \frac{1}{2}
    \sum_{\elementsymbol \in \bigrankorderingalpha{\infty}}
    \left(
    1
    -
    \delta_{\probsymbol_{\elementsymbol,\indexaraw},\probsymbol_{\elementsymbol,\indexbraw}}
    \right)
    \textnormal{max}
    \left(
    \probsymbol_{\elementsymbol,\indexaraw},\probsymbol_{\elementsymbol,\indexbraw}
    \right),
    \label{eq:probturbdiv.elementprobturbdiv_inflimit}
  \end{equation}
\item
  where the normalization
  from Eq.~\ref{eq:probturbdiv.probdiv_norm}
  has become
\item
  \begin{gather}
    \ptdnormalpha{\infty}
    =
    \sum_{\elementsymbol \in \bigrankorderingalpha{\infty}}
    \Bigm(
    \probsymbol_{\elementsymbol,\indexaraw}
    +
    \probsymbol_{\elementsymbol,\indexbraw}
    \Bigm)
    =
    1 + 1
    =
    2.
    \label{eq:probturbdiv.elementprobturbdiv_infinitylimit_norm}
  \end{gather}
\item
  The dominant contributions to probability-turbulence divergence in the
  $\alpha \rightarrow \infty$ limit
  therefore
  come from the most common types in each system, providing they are not equally abundant.
\end{textblock}

\revisionnote{
  We have swapped the order of following two sections.
  Connections between PTD and other measures now come first.
  We follow by empirical examples showing examples for special values of $\alpha$.
}

\section{Connections to other divergences and entropies}
\label{sec:probturbdiv.links}

\begin{revisionbar}
  
\subsection{Links to existing probability-based divergences}

\revisionnote{
  This subsection has been improved and expanded.
}

\begin{textblock}
\item
  Probability-turbulence divergence shares characteristics
  with and generalizes many other divergences
\item
  (see Refs.~\cite{deza2006a} and \cite{cha2007a} for two
  example compendia).
\item 
  In particular, we find known distances and similarities
  which correspond with or function similarly to 
\item
  probability-turbulence divergence
  for $\alpha$=0, 1/2, 1, and $\infty$.
\end{textblock}

\end{revisionbar}

\begin{revisionbar}

\subsubsection{Case 1: $\mathbf{\alpha = 0}$}
\label{subsubsec:probturbdiv.alpha=0}

\begin{textblock}
\item
  For $\alpha$=$0$,
  probability-turbulence divergence
  partners 
  the similarity measure S{\o}rensen-Dice coefficient,
\item 
  $
  S^{\textnormal{S-D}}(
  \bigprobsymbol_{\indexaraw}
  \,\Vert\,
  \bigprobsymbol_{\indexbraw}
  )
  $~\cite{dice1945a,sorensen1948a,looman1960a},
\item 
  which was independently developed in the context of ecology
  by
  Dice (1945) and S{\o}rensen (1948)
  (see also Ref.~\cite{stigler1980a}).
\item 
  For two systems, the S{\o}rensen-Dice coefficient
  is the number of shared types relative to the mean of the
  number of types in each system.
\item 
  Using our notation, and referring
  back to~Eq.~\ref{eq:probturbdiv.probdiv_alpha_zero_exclusive},
  we have:
\item 
  \begin{align}
    S^{\textnormal{S-D}}(
    \bigprobsymbol_{\indexaraw}
    \,\Vert\,
    \bigprobsymbol_{\indexbraw}
    )
    &
    =
    \frac{2}{
      \left(
      \Ntypesa + \Ntypesb
      \right)
    }
    \sum_{\elementsymbol \in \bigrankorderingalpha{0}}
    \left(
    1
    -
    \delta_{\probsymbol_{\elementsymbol,\indexaraw},0}
    -
    \delta_{0,\probsymbol_{\elementsymbol,\indexbraw}}
    \right)
    \nonumber
    \\
    &
    =
    1
    -
    \probdiv{0}(
    \bigprobsymbol_{\indexaraw}
    \,\Vert\,
    \bigprobsymbol_{\indexbraw}
    ),
  \end{align}
\item
  where we are again summing over the union of types
  $\bigrankorderingalpha{0}$.
\item
  The quantity
  $
  \left(
  1
  -
  \delta_{\probsymbol_{\elementsymbol,\indexaraw},0}
  -
  \delta_{\probsymbol_{\elementsymbol,\indexbraw},0}
  \right)
  $
\item
  is 1 when a type appears in both systems
  and 0 otherwise.
\end{textblock}

\begin{textblock}
\item
  The S{\o}rensen-Dice coefficient has arisen
  in many settings, with different names.
\item 
  For example, in statistics, the S{\o}rensen-Dice coefficient
  is the $F_{1}$ score of a test's accuracy~\cite{vanrijsbergen1979a,sasaki2007a}.
\end{textblock}

\end{revisionbar}

\begin{revisionbar}

\subsubsection{Case 2: $\mathbf{\alpha = 1/2}$}
\label{subsubsec:probturbdiv.alpha=1/2}

\begin{textblock}
\item
  Examples of divergences matching the internal
  structure of $\probdiv{1/2}$
  include
  the Hellinger distance~\cite{hellinger1909a} (1909),
  $\strongdistance^{\textnormal{Hel}}$,
  the proportional Matusita distance~\cite{matusita1955a} (1955),
  $\strongdistance^{\textnormal{Mat}}$,
  and
  the squared-chord distance~\cite{gavin2003a} (2003),
  $\strongdistance^{\textnormal{sqd-chd}}$:
\item
  \begin{gather}
    \left[
      \probdiv{1/2}(
      \bigprobsymbol_{\indexaraw}
      \,\Vert\,
      \bigprobsymbol_{\indexbraw}
      )
      \right]^{3/4}
    \\
    \nonumber
    \propto
    2^{-1/2}
    \strongdistance^{\textnormal{Hel}}(
      \bigprobsymbol_{\indexaraw}
      \,\Vert\,
      \bigprobsymbol_{\indexbraw}
      )
    \\
    \nonumber
    =
    \strongdistance^{\textnormal{Mat}}(
      \bigprobsymbol_{\indexaraw}
      \,\Vert\,
      \bigprobsymbol_{\indexbraw}
      )
    \\
    \nonumber
    =
    \left[
    \strongdistance^{\textnormal{\,sqd-chd}}(
      \bigprobsymbol_{\indexaraw}
      \,\Vert\,
      \bigprobsymbol_{\indexbraw}
      )
    \right]^{1/2}
    \nonumber
    \\
    =
    \left[
      \sum_{\elementsymbol \in \bigrankorderingalpha{1/2}}
      \Bigm(
      \left[\,\probsymbol_{\elementsymbol,\indexaraw}\right]^{1/2}
      -
      \left[\,\probsymbol_{\elementsymbol,\indexbraw}\right]^{1/2}
      \Bigm)^{2}
      \right]^{1/2}.
    \label{eq:probturbdiv.hellinger-matusita}    
  \end{gather}
\end{textblock}

\end{revisionbar}

\begin{revisionbar}

\subsubsection{Case 3: $\mathbf{\alpha = 1}$}
\label{subsubsec:probturbdiv.alpha=1}

\revisionnote{
  Section expanded to explain the $L_{q}$ distance
  match with PTD in terms of type contributions.
}

\revisionnote{
  We have adjusted notation from standard $L_{p}$ to $L_{q}$
  to avoid double use of $p$.
}

\begin{textblock}
\item
  In terms of the probability-turbulence divergence's internal structure of
\item
  $
  \lvert
  \left[\,\probsymbol_{\elementsymbol,\indexaraw}\right]^{\alpha}
  -
  \left[\,\probsymbol_{\elementsymbol,\indexbraw}\right]^{\alpha}
  \lvert
  $,
\item
  a large selection of divergences match up with
  the $\alpha$=1 instance.
\item
  These include all constructions built
  around the $L_{q}$ space
  Minkowski distance~\cite{minkowski1896a,newman1973a,krause1987a,cha2007a}:
\item
  \begin{equation}
    \strongdistance_{q}(
    \bigprobsymbol_{\indexaraw}
    \,\Vert\,
    \bigprobsymbol_{\indexbraw}
    )
    =
    \left[
      \sum_{\elementsymbol \in \bigrankorderingalpha{1}}
      \Bigm\lvert
      \left[\,\probsymbol_{\elementsymbol,\indexaraw}\right]^{1}
      -
      \left[\,\probsymbol_{\elementsymbol,\indexbraw}\right]^{1}
      \Bigm\lvert^{q}
      \right]^{1/q}.
    \label{eq:probturbdiv.minkowski}
  \end{equation}
\item
  We have explicitly included the powers of 1
  for the probabilities
  to emphasize
  the match with $\alpha$=1 for PTD.
\end{textblock}

\begin{textblock}
\item
  For example, for $q=1$, we have
  the city block (or Manhattan) distance:
\item
  \begin{equation}
    \strongdistance_{1}(
    \bigprobsymbol_{\indexaraw}
    \,\Vert\,
    \bigprobsymbol_{\indexbraw}
    )
    =
    \sum_{\elementsymbol \in \bigrankorderingalpha{1}}
    \Bigm\lvert
    \left[\,\probsymbol_{\elementsymbol,\indexaraw}\right]^{1}
    -
    \left[\,\probsymbol_{\elementsymbol,\indexbraw}\right]^{1}
    \Bigm\lvert,
    \label{eq:probturbdiv.cityblock}
  \end{equation}
\item
  and for $q=2$, 
  the standard spatial (or Euclidean) distance:
\item
  \begin{equation}
    \strongdistance_{2}(
    \bigprobsymbol_{\indexaraw}
    \,\Vert\,
    \bigprobsymbol_{\indexbraw}
    )
    =
    \left[
      \sum_{\elementsymbol \in \bigrankorderingalpha{1}}
      \Bigm\lvert
      \left[\,\probsymbol_{\elementsymbol,\indexaraw}\right]^{1}
      -
      \left[\,\probsymbol_{\elementsymbol,\indexbraw}\right]^{1}
      \Bigm\lvert^{2}
      \right]^{1/2}.
    \label{eq:probturbdiv.standardistance}
  \end{equation}

\end{textblock}

\begin{textblock}
\item
  Note that in the
  $q \rightarrow 0$
  and
  $q \rightarrow \infty$ limits,
  the connection between the $L_{q}$ distance
  and $\probdiv{1}$ breaks down.
\item 
  In general, as $q \rightarrow 0$,
  for two vectors
  $\vec{z}_{1}$
  and
  $\vec{z}_{2}$,
  $L_{q}(\vec{z}_{1},\vec{z}_{2})$ will be:
  0 if 
  $\vec{z}_{1} = \vec{z}_{2}$;
  1 if the vectors differ in exactly one coordinate;
  and
  $\infty$ otherwise.
\item
  Because two probability distributions cannot differ in just one coordinate,
  $
  \strongdistance_{0}(
  \bigprobsymbol_{\indexaraw}
  \,\Vert\,
  \bigprobsymbol_{\indexbraw}
  )
  $
  is either 0 or $\infty$.
\item
  And because of how we have constructed $\probdiv{\alpha}$ to behave
  well and meaningfully in the limit $\alpha \rightarrow 0$,
  the link breaks down.
\end{textblock}

\begin{textblock}
\item
  For the limit $\alpha \rightarrow \infty$,
  $
  \strongdistance_{\infty}(
  \bigprobsymbol_{\indexaraw}
  \,\Vert\,
  \bigprobsymbol_{\indexbraw}
  )
  =
  \max_{\elementsymbol}
  \lvert
  \probsymbol_{\elementsymbol,\indexaraw}
  -
  \probsymbol_{\elementsymbol,\indexbraw}
  \lvert
  $
  which is the 
  Chebyshev distance~\cite{deza1996a}.
\item
  For $\probdiv{\infty}$, the sum remains and maxima
  are determined for each element rather than overall
  (Eq.~\ref{eq:probturbdiv.elementprobturbdiv_inflimit}).
\end{textblock}

\begin{textblock}
\item
  Now, while the overall values for these
  $L_{q}$ distances
  will vary,
\item
  the rank orderings of their contributing types
  will all be identical to that of $\probdiv{1}$
  providing $0 < q < \infty$.
\item
  To see this,
  we first leave aside
  the overall power of $1/q$
  for
  $
  \strongdistance_{q}(
  \bigprobsymbol_{\indexaraw}
  \,\Vert\,
  \bigprobsymbol_{\indexbraw}
  )
  $, 
  as well as prefactors
  for $\probdiv{1}$ (see Eq.~\ref{eq:probturbdiv.probdiv_definition}).
\item
  For type $\elementsymbol$,
  we then have the summands
  $
  \lvert
  \probsymbol_{\elementsymbol,\indexaraw}
  -
  \probsymbol_{\elementsymbol,\indexbraw}
  \lvert^{q}
  $
  for
  $
  \strongdistance_{q}(
  \bigprobsymbol_{\indexaraw}
  \,\Vert\,
  \bigprobsymbol_{\indexbraw}
  )
  $, 
  and
  $
  \lvert
  \probsymbol_{\elementsymbol,\indexaraw}
  -
  \probsymbol_{\elementsymbol,\indexbraw}
  \lvert^{1/2}
  $
  for 
  $\probdiv{1}(
  \bigprobsymbol_{\indexaraw}
  \,\Vert\,
  \bigprobsymbol_{\indexbraw}
  ).
  $
\item
  Because
  if $\beta > 0$,
  $
  \lvert
  z_{1}
  \lvert^{\beta}
  <
  \lvert
  z_{2}
  \lvert^{\beta}
  $
  for all
  $
  \lvert
  z_{1}
  \lvert
  <
  \lvert
  z_{2}
  \lvert,
  $
  the orderings of types by contribution
  will be the same
  (i.e., the powers $q$ and $1/2$ do not matter, just that they are both
  between 0 and $\infty$).
\end{textblock}

\end{revisionbar}

\begin{revisionbar}

\subsubsection{Case 4: $\mathbf{\alpha = \infty}$}
\label{subsubsec:probturbdiv.alpha=infty}

\begin{textblock}
\item
  Finally, in the $\alpha$=$\infty$ limit,
\item 
  $\probdiv{\infty}$ is akin to,
  and in practice the same as,
\item 
  the Motyka  distance~\cite{deza2006a},
  $\strongdistance^{\textnormal{Mot}}$:
\item
  \begin{equation}
    \strongdistance^{\textnormal{Mot}}
    =
    \frac{
      \sum_{\elementsymbol \in \bigrankorderingalpha{\infty}}
      \max_{\elementsymbol}
      (
      \probsymbol_{\elementsymbol,\indexaraw},
      \probsymbol_{\elementsymbol,\indexbraw}
      )
    }{
      \sum_{\elementsymbol \in \bigrankorderingalpha{\infty}}
      (
      \probsymbol_{\elementsymbol,\indexaraw}
      +
      \probsymbol_{\elementsymbol,\indexbraw}
      )
    }.
    \label{eq:probturbdiv.motykadistance}    
  \end{equation}
\item
  Because of the factor
  $\delta_{\probsymbol_{\elementsymbol,\indexaraw},\probsymbol_{\elementsymbol,\indexbraw}}$
  in Eq.~\ref{eq:probturbdiv.elementprobturbdiv_inflimit},
  if there are any types which have the same probability in both systems,
  then there will be disagreement.
\item
  However, for real, large-scale systems, the total number of types will
  likely differ, greatly reducing the chance of any equal probabilities.
\item
  And where the type set is fixed for both systems, large sample numbers should
  prevent equivalences.
\end{textblock}

\begin{textblock}
\item
  We summarize the connections between
  probability-turbulence divergence
  and other measures in Tab.~\ref{tab:probturbdiv.connections}.
\end{textblock}

\end{revisionbar}

\renewcommand\floatpagefraction{0.5}

\begin{table}
  \renewcommand{\arraystretch}{1.8}
  \rowcolors{2}{gray!5}{gray!15}
  \begin{tabular}{cp{8pt}>{\raggedright\arraybackslash}p{0.9\columnwidth}p{4pt}}
    \rowcolor{gray!25}
    \hline
    ~$\mathbf{\alpha}$~
    &
    &
    \textbf{Measures with matching type orderings}
    &
    \\
    \hline
    $0$
    &
    &
    Similarity measure S{\o}rensen-Dice coefficient~\cite{dice1945a,sorensen1948a,looman1960a},
    $F_{1}$ score of a test's accuracy~\cite{vanrijsbergen1979a,sasaki2007a}
    &
    \\
    $1/2$
    &
    &
    Hellinger distance~\cite{hellinger1909a},
    Matusita distance~\cite{matusita1955a},
    squared-chord distance~\cite{gavin2003a}
    &
    \\
    $1$
    &
    &
    Many including all $L_{q}$ distance constructions
    for $0 < q < \infty$~\cite{cha2007a}
    &
    \\
    $\infty$
    &
    &
    Motyka distance~\cite{deza2006a}
    (providing no types have the exact same probabilities)
    &
    \\
    \hline
  \end{tabular}
  \caption{
    Correspondences between probability-turbulence divergence and
    extant difference measures for distributions.
    Equivalence is at the level of the contribution by individual types
    to overall score.
  }
  \label{tab:probturbdiv.connections}
  \renewcommand{\arraystretch}{1}
\end{table}

\subsection{Generalized entropies and Hill numbers}
\label{subsec:probturbdiv.entropy}

\begin{textblock}
\item
  While none of these other divergences provide direct tunability of
  the type probability---a severe limitation, as we intend our examples
  below in Sec.~\ref{sec:probturbdiv.graphicalinstrument}
  will convey---there
  are well-established quantities which do.
\end{textblock}

\begin{textblock}
\item
  As we observed for rank-turbulence divergence in~\cite{dodds2023a},
  the parameter $\alpha$'s effect is similar to that of its counterparts in
  various kinds of generalized entropy
\item
  including those of R{\'e}nyi and Tsallis~\cite{renyi1961a,tsallis2001a,keylock2005a}
\item 
  and, more directly,
\item 
  the diversity indices (or Hill numbers) from ecology~\cite{hill1973a,jost2006a}.
\item 
  We make some links with R{\'e}nyi entropy,
  leaving other connections for possible future work.
\end{textblock}

\begin{textblock}
\item
  R{\'e}nyi entropy,
  ${}^{\alpha\!}H$,
\item 
  and the associated diversity index,
\item 
  ${}^{\alpha\!}N$,
  are defined as:
\item 
  \begin{equation}
    {}^{\alpha\!}H
    =
    \textnormal{ln}
    \,
    \left[
      {}^{\alpha\!}N
      \right]
    =
    \frac{1}
         {1-\alpha}
         \textnormal{ln}
         \,
         \left(
         \sum_{\elementsymbol}
         \left[\,\probsymbol_{\elementsymbol}\right]^{\alpha}
         \right),
         \label{eq:probturbdiv.entropy}
  \end{equation}
\item
  and
\item
  \begin{equation}
    {}^{\alpha\!}N
    =
    \left(
    \sum_{\elementsymbol}
    \left[\,\probsymbol_{\elementsymbol}\right]^{\alpha}
    \right)^{
      1/(1-\alpha)
    },
    \label{eq:probturbdiv.hillnumber}
  \end{equation}
\item
  where $\alpha \ge 0$.
  We acknowledge that, at the risk of a minor dislocation from
  relevant literature,
\item
  we have had to confront some notation peril here
  as a standard notation for the diversity index
  is ${}^{\alpha\!}D$.
\item
  We have also already used $N$ in our present paper
  but this choice tracks sensibly:
\item
  As $\alpha \rightarrow 0$, we retrieve the natural logarithm
  of the number of distinct types $N$ (species richness in ecology)
  for R{\'e}nyi entropy,
\item
  and therefore the diversity index is ${}^{0\!}N = N$.
\item 
  As $\alpha \rightarrow \infty$, the most abundant type will dominate,
\item
  with min-entropy the limit:
\item
  $
  {}^{\infty\!}N
  =
  \textnormal{min}_{\elementsymbol}
  1/\probsymbol_{\elementsymbol}
  =
  1/\textnormal{max}_{\elementsymbol}
  \probsymbol_{\elementsymbol}.
  $
\item
  In the $\alpha \rightarrow 1$ limit,
\item 
  we recover Shannon's entropy, $H$,
\item 
  as well as
\item 
  $
  {}^{1\!}N
  =
  e^{H_{1}}
  =
  e^{H}.
  $
\end{textblock}

\begin{textblock}
\item
  There are similar aspects for probability-turbulence divergence
  and the diversity index in the limits of $\alpha$=0 and $\infty$.
\item 
  For $\alpha$=0,
  for example,
  both reduce to quantities involving simple
  counts of distinct types.
\item 
  Nevertheless,
  we note that we cannot construct probability-turbulence divergence
  from manipulations of 
  R{\'e}nyi entropy or the diversity index.
\item 
  We can, roughly speaking, only create a difference of sums
  whereas we need a sum of absolute differences with suitable exponents.
\end{textblock}

\section{Probability-Turbulence Divergence Allotaxonographs}
\label{sec:probturbdiv.graphicalinstrument}

\revisionnote{
  Jane Austen section reconfigured to show more values of $\alpha$
  and to present them in order.
}

\revisionnote{
  We now emphasize the importance of all values of $\alpha$.
  We also have changed from writing `optimal' to `scale-equalizing' $\alpha$.
}

\begin{figure*}[t!]
  \includegraphics[width=1.1\textwidth,center]{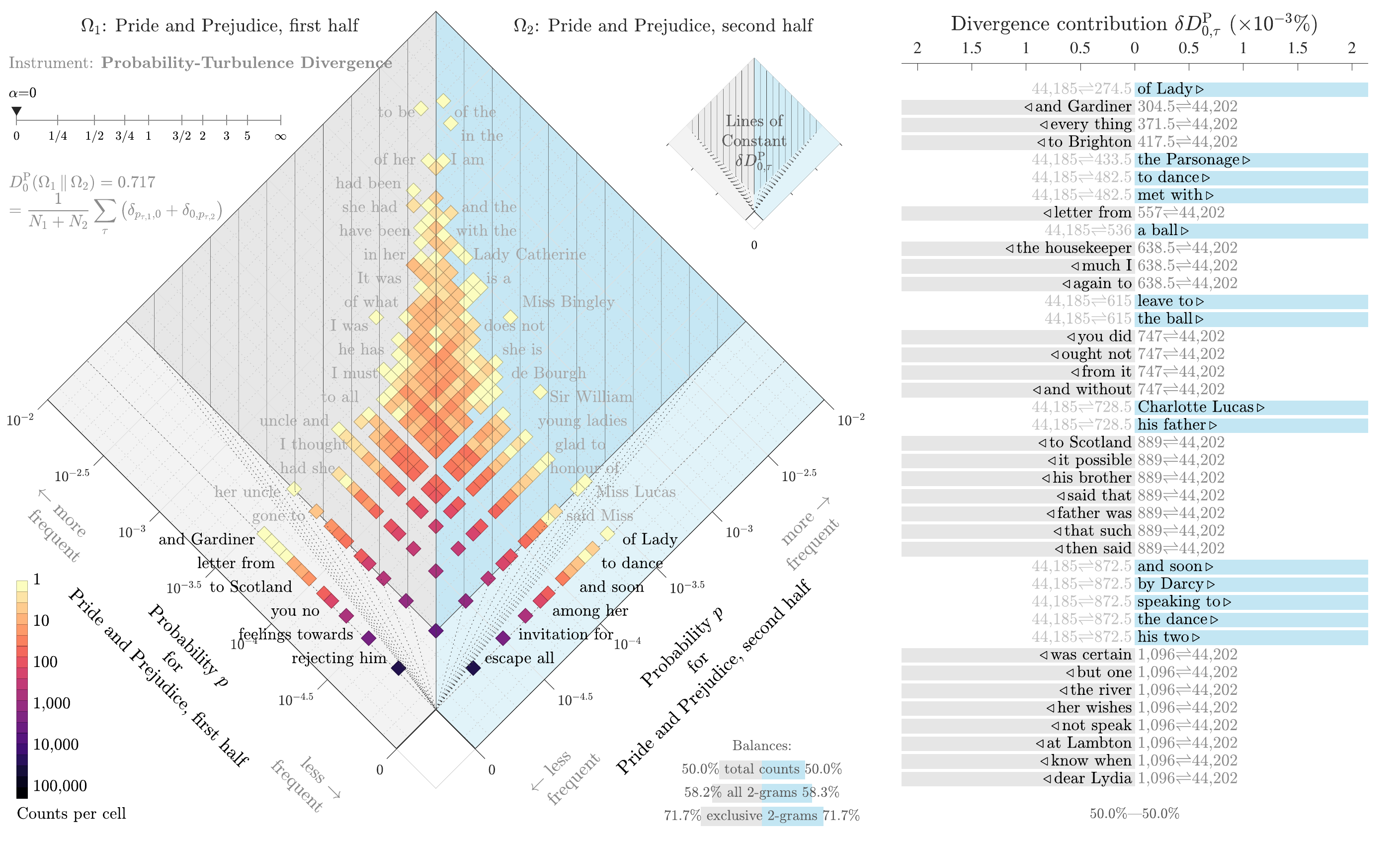}
  \caption{
    \revisionnote{Figure re-ordered and caption built of out of earlier versions.}\\
    \revision{
    \textbf{Allotaxonograph comparing 2-gram usage in the first and second halves
    of Jane Austen's Pride and Prejudice
    using probability-turbulence divergence with $\alpha$=$0$,
    $\probdiv{0}$.}
    To create our allotaxonographs, we bin all non-zero probability pairs
    $(\log_{10}\probsymbol_{\elementsymbol,\indexaraw},\log_{10}\probsymbol_{\elementsymbol,\indexbraw})$
    in logarithmic space.
    Colors indicate counts of 2-grams per cell,
    and we highlight example 2-grams along the edges of the histogram.
    For pairs where one of the probabilities is zero,
    we add a separate rectangular panel along the bottom of each axis (lighter gray and lighter blue).
    Contour lines indicate where probability-turbulence divergence is constant
    (the jump to the zero probability region necessitates a break in smoothness).
    The gray scale for 2-grams is indexed by their percentage
    contribution to probability-turbulence divergence,
    $\probdivelement{0}$.
    \textbf{Ranked list on the right:}
    We order the most salient 2-grams according to 
    their overall contribution
    $\probdivelement{0}$
    which we mark by bar length.
    We show the rank pair for each 2-gram in light gray opposite each 2-gram.
    \textbf{For $\alpha$=$0$ at one extreme of the parameter's range,
    probability-turbulence divergence
    elevates exclusive types above all
    types that appear in both systems,
    and the ranked list on the right comprises
    only system-exclusive 2-grams.}
    Per Sec.~\ref{subsec:probturbdiv.probturbdivzero}
    and
    Eq.~\ref{eq:probturbdiv.probdiv_alpha_zero_exclusive},
    exclusive types each equally contribute
    $
    \frac{1}{
      \left(
      \Ntypesa + \Ntypesb
      \right)
    }
    $
    to
    $\probdiv{0}$
    while types appearing in both systems
    have zero weight.
    The ordering of 2-grams is determined by maintaining
    their contribution order as $\alpha$ approaches 0.
    We force the contour lines in the main body of the histogram
    to remain equally spaced,
    even as they all represent 0 in the $\alpha \rightarrow 0$ limit.
    Outside of the main body of the histogram, all contour lines
    travel to the $(0,0)$ point (which is $(-\infty,-\infty)$ in log-space).
    See Sec.~\ref{sec:probturbdiv.links} 
    for the connection between
    $\probdiv{0}$
    and the
    S{\o}rensen-Dice coefficient~\cite{dice1945a,sorensen1948a,looman1960a}
    and
    the $F_{1}$ score~\cite{vanrijsbergen1979a,sasaki2007a}.
    See also Flipbooks
    \href{\zenodofileslink/allotaxonometer-flipbook-1-probability-divergence-pride-and-prejudice-1-grams.pdf?download=1}{S1},
    \href{\zenodofileslink/allotaxonometer-flipbook-2-probability-divergence-pride-and-prejudice-2-grams.pdf?download=1}{S2},
    and
    \href{\zenodofileslink/allotaxonometer-flipbook-3-probability-divergence-pride-and-prejudice-3-grams.pdf?download=1}{S3}
    in the supplementary material,
    per Sec.~\ref{sec:probturbdiv.methods.suppmaterial}.
    }
  }
  \label{fig:probturbdiv.allotaxonometer9000-jane-austen-2-grams-0}
\end{figure*}

\begin{figure*}[t!]
  \includegraphics[width=1.1\textwidth,center]{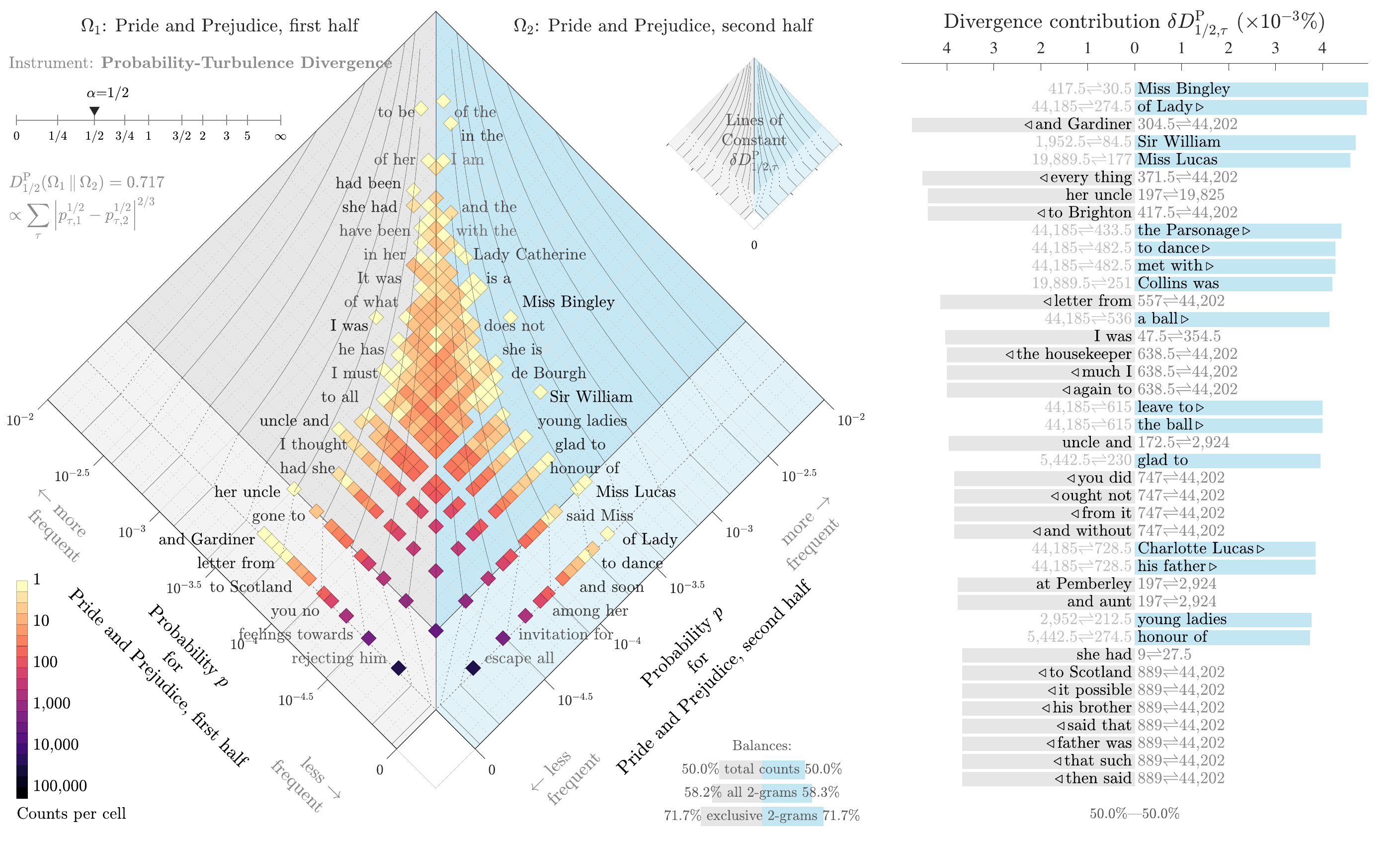}
  \caption{
    \revisionnote{New figure}\\
    \textbf{Allotaxonograph comparing 2-gram usage in the first and second halves
      of Jane Austen's Pride and Prejudice
      using probability-turbulence divergence with $\alpha$=1/2,
      $\probdiv{1/2}$.}
  }
  \label{fig:probturbdiv.allotaxonometer9000-jane-austen-2-grams-1/2}
\end{figure*}

\begin{figure*}[t!]
  \includegraphics[width=1.1\textwidth,center]{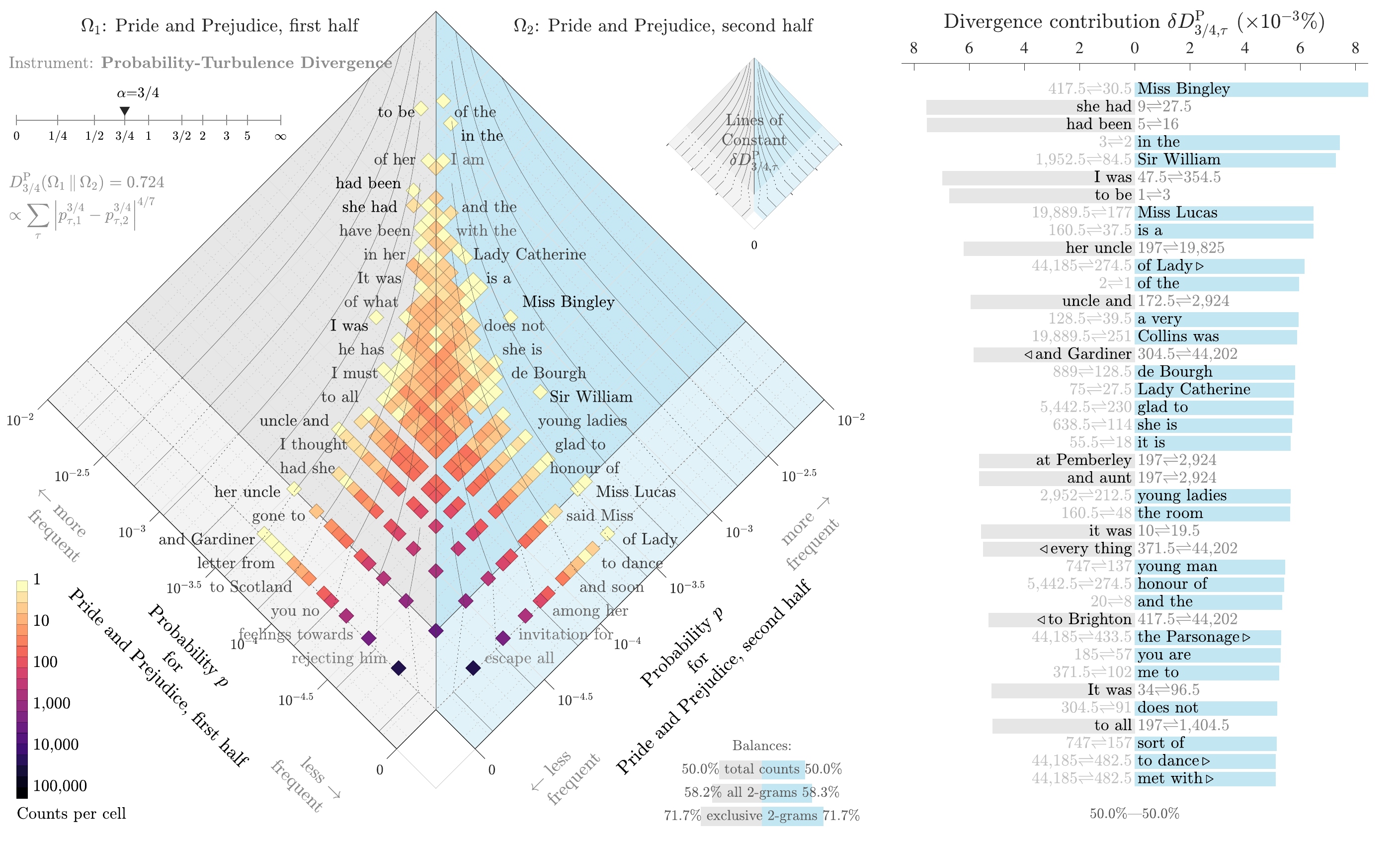}
  \caption{
    \revisionnote{Figure reordered}\\
    \textbf{Allotaxonograph comparing 2-gram usage in the first and second halves
      of Jane Austen's Pride and Prejudice
      using probability-turbulence divergence with
      a scale-equalizing $\alpha$=3/4,
      $\probdiv{3/4}$.}
    By scale-equalizing, we mean that the contour lines reasonably fit with the outer shape of the
    histogram, thereby giving the list on the right a mixture of contributions from rare to common 2-grams.
  }
  \label{fig:probturbdiv.allotaxonometer9000-jane-austen-2-grams-3/4}
\end{figure*}

\begin{figure*}[t!]
  \includegraphics[width=1.1\textwidth,center]{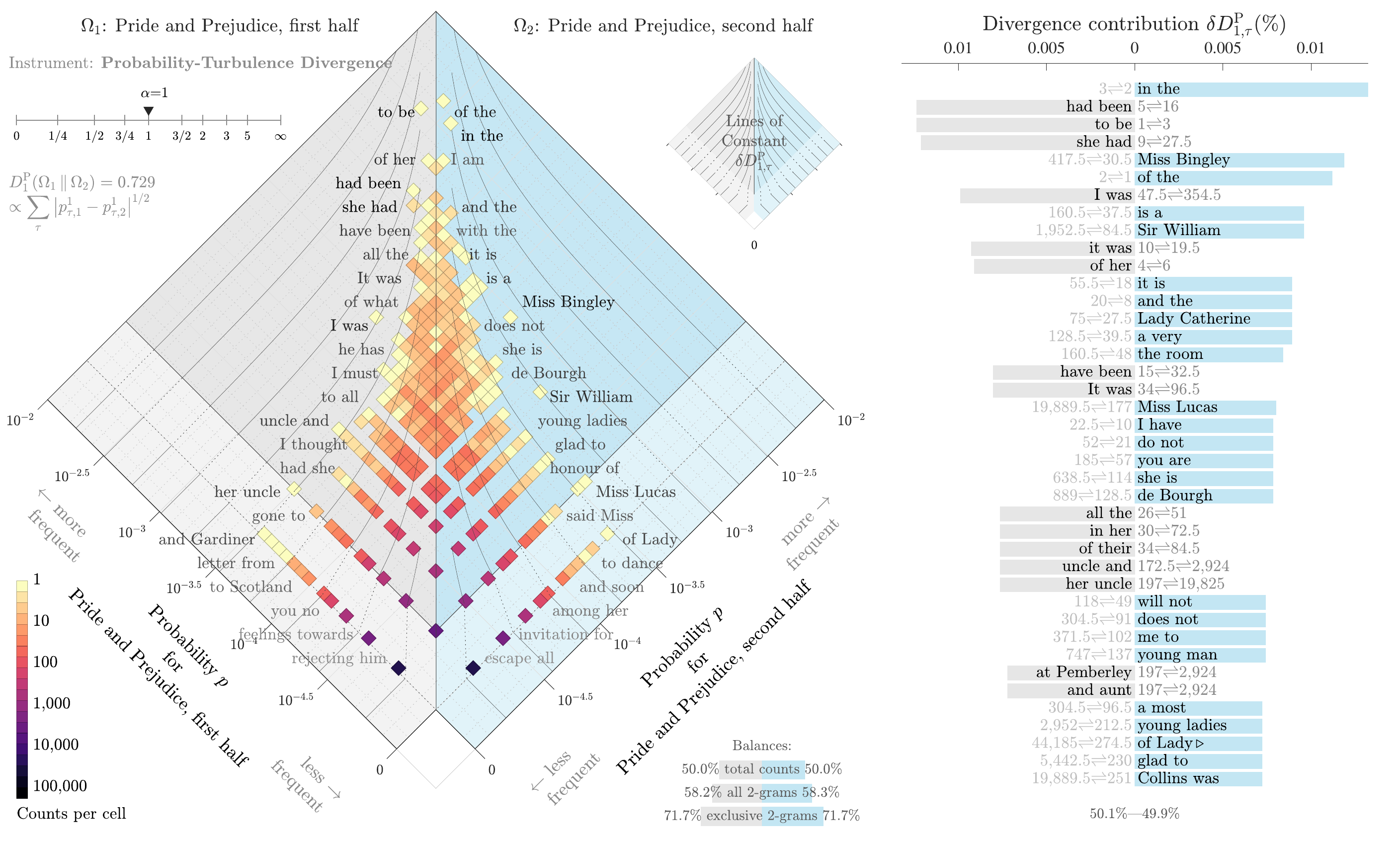}
  \caption{
    \revisionnote{New figure}\\
    \textbf{Allotaxonograph comparing 2-gram usage in the first and second halves
      of Jane Austen's Pride and Prejudice
      using probability-turbulence divergence with $\alpha$=1,
      $\probdiv{1}$.}
  }
  \label{fig:probturbdiv.allotaxonometer9000-jane-austen-2-grams-1}
\end{figure*}

\begin{figure*}[t!]
  \includegraphics[width=1.1\textwidth,center]{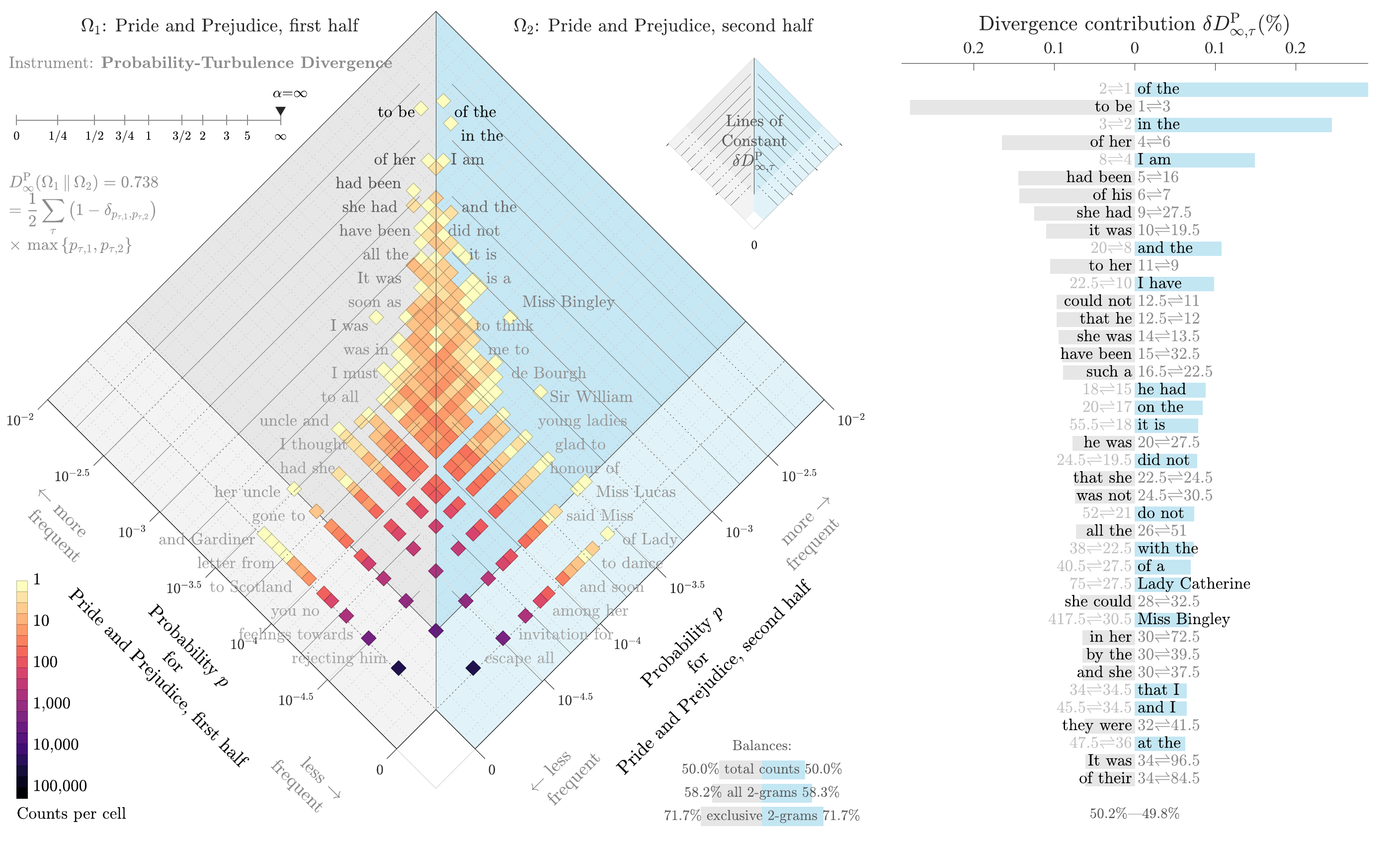}
  \caption{
    \revisionnote{Figure in new order}\\
    \textbf{Allotaxonograph comparing 2-gram usage in the first and second halves
      of Jane Austen's Pride and Prejudice
      using probability-turbulence divergence with $\alpha$=$\infty$,
      $\probdiv{\infty}$.}
    The most common 2-grams now fully dominate the contributions 
    as indicated in the ranked list and by the darker shading of the
    2-grams at the top of the histogram.
    As for $\alpha$=$0$ in
    Fig.~\ref{fig:probturbdiv.allotaxonometer9000-jane-austen-2-grams-0},
    the contours do not conform to the edges of the histogram.
    The dominant 2-grams for $\alpha = \infty$
    largely comprise function words with the exception of
    `Lady Catherine'
    and
    `Miss Bingley'
    (characters most commonly referred to with a 2-gram).
  }
  \label{fig:probturbdiv.allotaxonometer9000-jane-austen-2-grams-infty}
\end{figure*}

\begin{figure*}[t!]
  \includegraphics[width=1.1\textwidth,center]{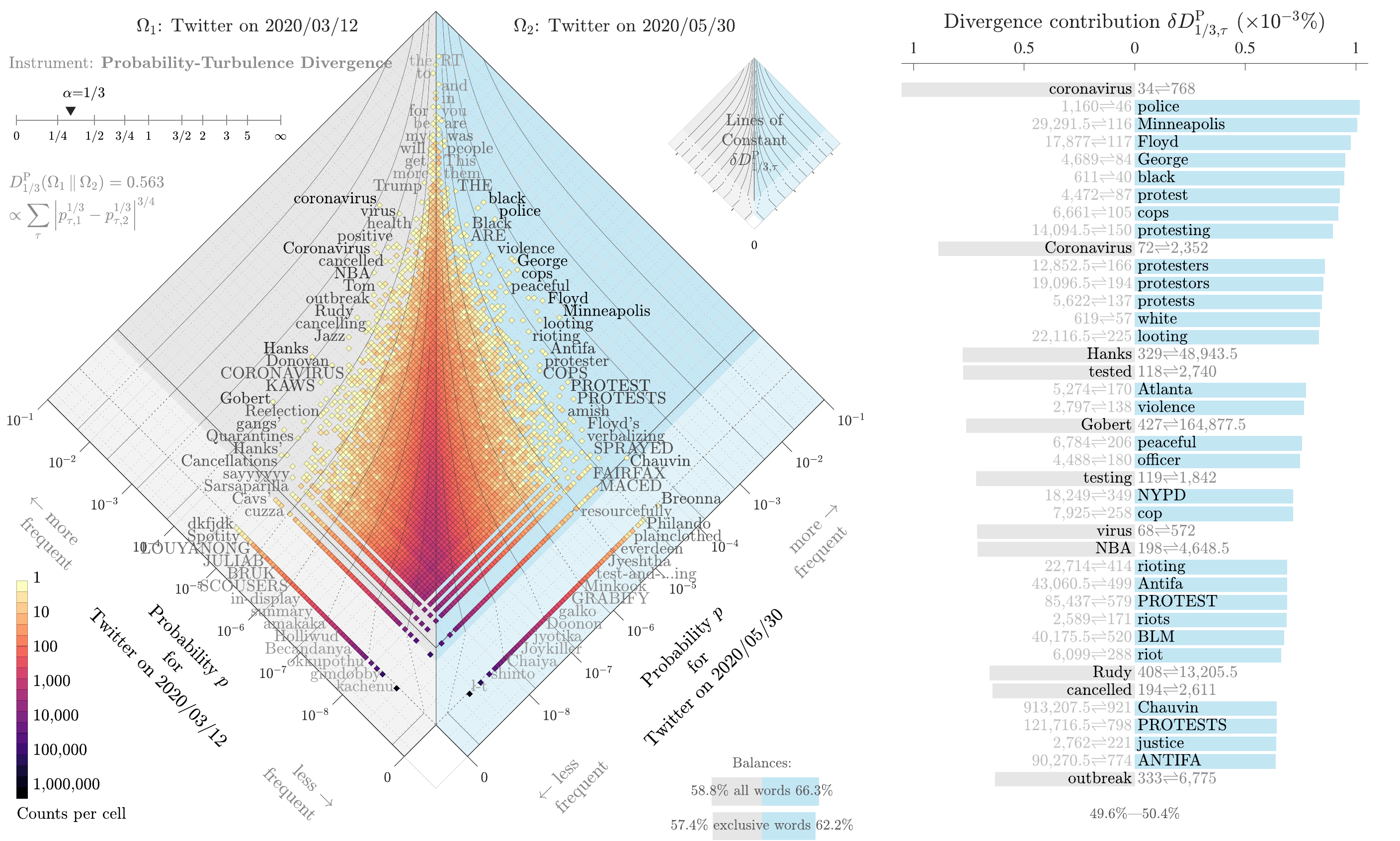}
  \caption{
    \textbf{Allotaxonograph using probability-turbulence divergence to
      compare normalized 1-gram usage rates
      on two days of Twitter, 2020/03/12 and 2020/05/30---key
      dates in the US for the COVID-19 pandemic and the Black Lives Matter protests following
      George Floyd's murder.}
    We assess $\alpha$=1/3 to be reasonably scale-equalizing for 1-grams.
    Details are the same
    as for Fig.~\ref{fig:probturbdiv.allotaxonometer9000-jane-austen-2-grams-0}.
    The days are according to Coordinated Universal Time (UTC)
    and the 1-grams are those containing Latin characters
    found in English-language tweets~\cite{alshaabi2021b,alshaabi2021c}.
    See Flipbook
    \href{\zenodofileslink/allotaxonometer-flipbook-4-probability-divergence-twitter-1-grams.pdf?download=1}{S4}
    for the instrument's variation as a function of $\alpha$.
  }
  \label{fig:probturbdiv.allotaxonometer9000-2020-03-12-2020-05-30-story-wrangler-1grams-all-prob-div}
\end{figure*}

\begin{figure*}[t!]
  \includegraphics[width=1.1\textwidth,center]{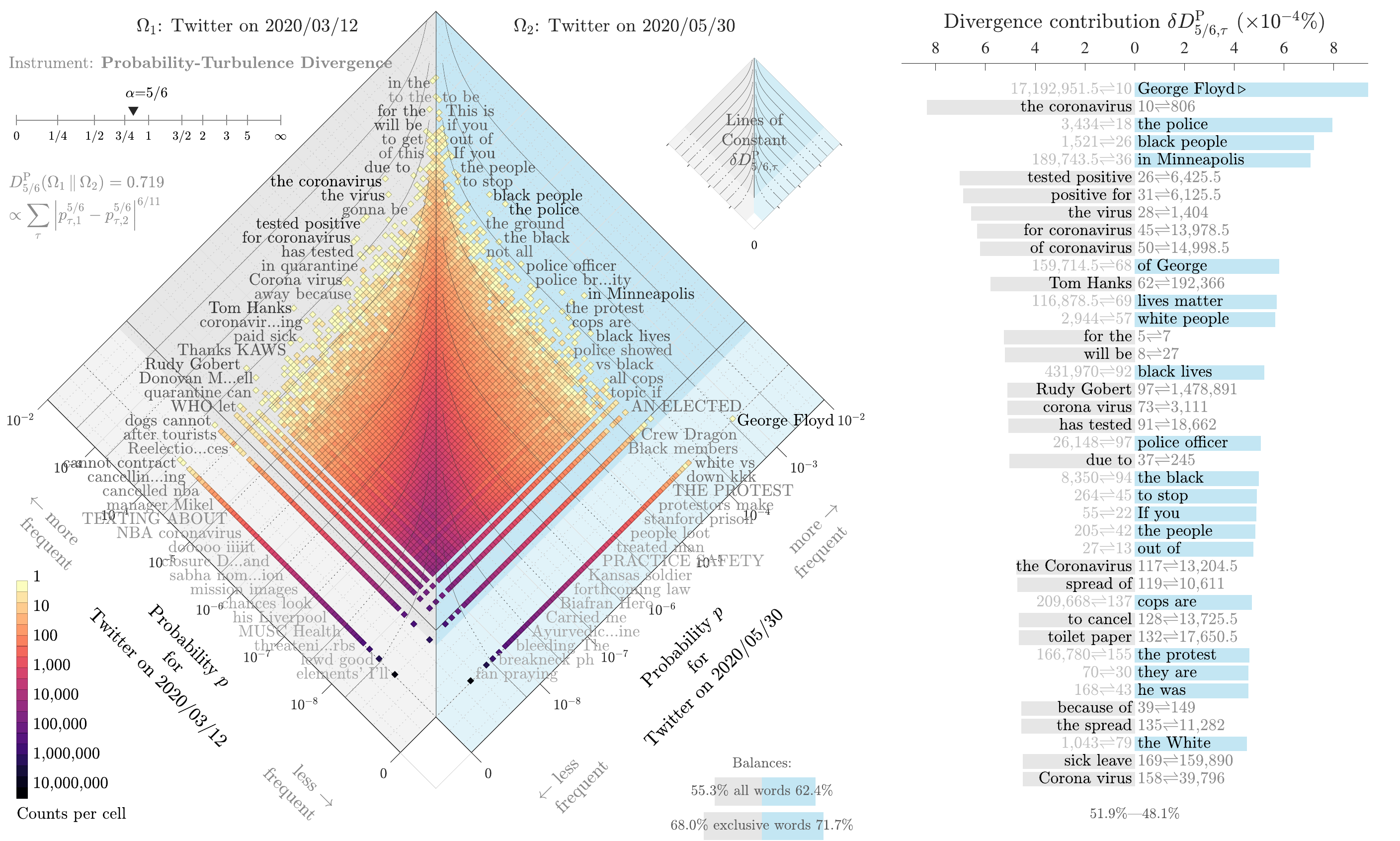}
  \caption{
    \textbf{Allotaxonograph using probability-turbulence divergence
      to compare normalized 2-gram usage ranks on two days of English-language Twitter, 2020/03/12 and 2020/05/30.}
    Details are the same
    as for Fig.~\ref{fig:probturbdiv.allotaxonometer9000-jane-austen-2-grams-0}
    and~\ref{fig:probturbdiv.allotaxonometer9000-2020-03-12-2020-05-30-story-wrangler-1grams-all-prob-div}.
    We see that the comparison of
    2-gram distributions produces a different, broader histogram 
    than that formed by 1-gram distributions
    (Fig.~\ref{fig:probturbdiv.allotaxonometer9000-2020-03-12-2020-05-30-story-wrangler-1grams-all-prob-div}).
    We choose $\alpha$=5/6 to provide a balance of 2-grams across five
    orders of magnitude for non-zero probability.
    In contrast to the 1-gram version,
    the top 2-grams are more evenly distributed on both sides of the list.
    While some 2-grams are function words combined with the 1-grams we
    saw in
    Fig.~\ref{fig:probturbdiv.allotaxonometer9000-2020-03-12-2020-05-30-story-wrangler-1grams-all-prob-div},
    meaningful 2-grams also appear (`Tom Hanks', `toilet paper', `George Floyd', and `police brutality').
    See Flipbook 
    \href{\zenodofileslink/allotaxonometer-flipbook-5-probability-divergence-twitter-2-grams.pdf?download=1}{S5}
    for the instrument's variation as a function of $\alpha$.
  }
  \label{fig:probturbdiv.allotaxonometer9000-2020-03-12-2020-05-30-story-wrangler-2grams-all-prob-div}
\end{figure*}

\begin{figure*}[t!]
  \includegraphics[width=1.1\textwidth,center]{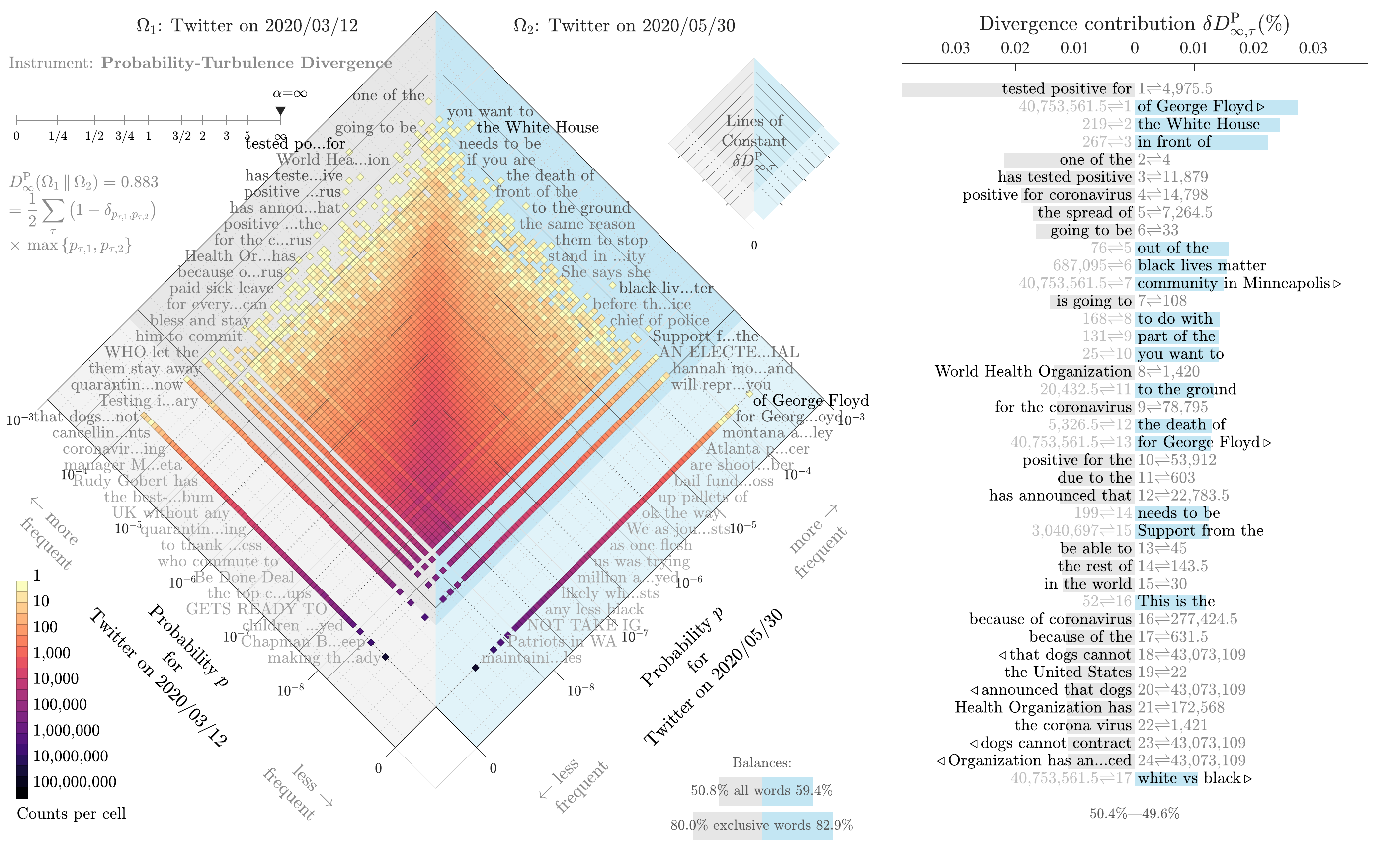}
  \caption{
    \textbf{Allotaxonograph using probability-turbulence divergence
      to compare 3-gram usage ranks on two days of English-language Twitter, 2020/03/12 and 2020/05/30.}
    Details are the same
    as for Fig.~\ref{fig:probturbdiv.allotaxonometer9000-jane-austen-2-grams-0}
    and~\ref{fig:probturbdiv.allotaxonometer9000-2020-03-12-2020-05-30-story-wrangler-1grams-all-prob-div}.
    The histogram has broadened even further out from the 2-gram, and now
    is well suited to probability-turbulence divergence
    with the extreme of $\alpha$=$\infty$, $\probdiv{\infty}$.
    Fragmentary and meaningful 3-grams appear alongside each other,
    including `World Health Organization'
    and
    `black lives matter'.
    Social amplification is also apparent as 3-grams for highly retweeted tweets
    dominate the rank list.
    See Flipbook
    \href{\zenodofileslink/allotaxonometer-flipbook-6-probability-divergence-twitter-3-grams.pdf?download=1}{S6}
    for the instrument's variation as a function of $\alpha$.
  }
  \label{fig:probturbdiv.allotaxonometer9000-2020-03-12-2020-05-30-story-wrangler-3grams-all-prob-div}
\end{figure*}

\begin{figure*}[t!]
  \includegraphics[width=1.1\textwidth,center]{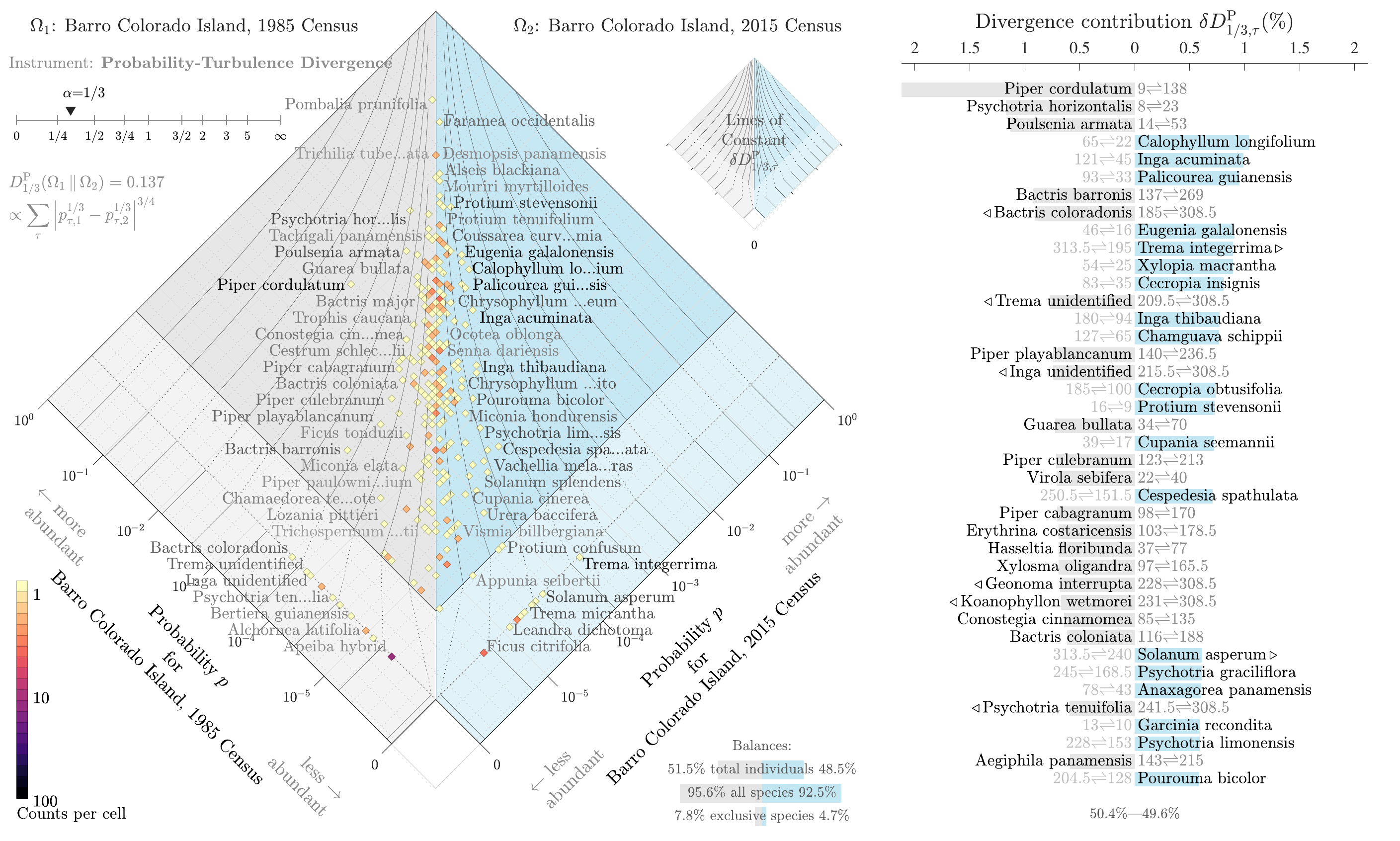}
  \caption{
    \textbf{Allotaxonograph using probability-turbulence divergence
    to compare tropical forest tree species abundance on Panama's Barro Colorado Island (BCI)
    for 5-year censuses completed in 1985 and 2015~\cite{condit2019a}.}
    The scale-equalizing
    choice of $\alpha$=1/3 produces a 
    set of dominant species
    reasonably well balanced across the abundance spectrum.
    See Ref.~\cite{dodds2023a} for the corresponding rank-turbulence divergence allotaxonographs.
    See Flipbook
    \href{\zenodofileslink/allotaxonometer-flipbook-7-probability-turbulence-divergence-barro-colorado-island.pdf?download=1}{S7}
    for the instrument's variation as a function of $\alpha$
    (Sec.~\ref{sec:probturbdiv.methods.suppmaterial}).
  }
  \label{fig:probturbdiv.allotaxonometer9000-1985-2015-barro-colorado002}
\end{figure*}

\begin{textblock}
\item
  We assert that the successful use of our rank- and probability-turbulence divergences
  is best achieved through consideration of rich graphical representations
  which we have called allotaxonographs~\cite{dodds2023a}.
\item 
  In this section, we present and describe allotaxonographs comparing probability
  distributions using probability-turbulence divergence for:
\end{textblock}
\begin{itemize}[noitemsep, topsep=0pt, label=\textbullet]
\item
  \revision{
  Normalized usage frequencies of 2-grams in the first and second halves
  of Jane Austen's Pride and Prejudice~\cite{austen2001a} for
  $\alpha$=0,
  1/2,
  3/4
  (scale-equalizing),
  1,
  and $\infty$
  (Figs.~\ref{fig:probturbdiv.allotaxonometer9000-jane-austen-2-grams-0}--\ref{fig:probturbdiv.allotaxonometer9000-jane-austen-2-grams-infty});
  }
\item
  Normalized usage frequencies of $n$-grams in all English-identified tweets on
  2020/03/12 and 2020/05/30 for $n$ = 1, 2, and 3 and for
  scale-equalizing values of
  $\alpha$ = 1/3,
  5/6,
  and $\infty$
  (Figs.~\ref{fig:probturbdiv.allotaxonometer9000-2020-03-12-2020-05-30-story-wrangler-1grams-all-prob-div},
  \ref{fig:probturbdiv.allotaxonometer9000-2020-03-12-2020-05-30-story-wrangler-2grams-all-prob-div},
  and
  \ref{fig:probturbdiv.allotaxonometer9000-2020-03-12-2020-05-30-story-wrangler-3grams-all-prob-div});
\item
  Relative abundances of tree species on Barro Colorado Island for five-year
  censuses concluding in 1985 and 2015
  for a scale-equalizing $\alpha$ = 1/3
  (Fig.~\ref{fig:probturbdiv.allotaxonometer9000-1985-2015-barro-colorado002}).
\end{itemize}
\begin{textblock}
\item
  \revision{
    In particular,
    the Pride and Prejudice examples for 2-grams will show how PTD may be adjusted to:
    emphasize rare $n$-grams ($\alpha$=0),
    emphasize the most common $n$-grams ($\alpha$=$\infty$),
    be scale-equalizing ($\alpha$=3/4),
    or behave like many extant distances and divergences
    ($\alpha$=1/2 and 1).
  }
\item 
  \revision{
    The choices of $\alpha$ for the three Twitter examples
    and the one from Barro Colorado Island
    further showcase how scale-equalizing fits are achieved
    by a range of values of $\alpha$.
  }
\item 
  There is no universal $\alpha$ that scale-equalizes
  turbulence between probability-rank distributions.
\end{textblock}

\begin{textblock}
\item
  The examples for 2-grams and 3-grams can also be seen as demonstrations of
  possible comparisons of features of complex networks and systems
  (e.g., 2-grams in text as directed edges).
\end{textblock}

\begin{textblock}
\item
  Based on our experience using RTD and PTD,
  we advise that either comparative instrument
  always be used to produce
  a sequence of allotaxonographs
  with the following 24 values of $\alpha$:
\item 
  \begin{equation}
    \vec{\alpha}
    =
    \left\{
      0,\,
      \frac{1}{12},\,
      \frac{2}{12},\,
      \ldots,\,
      \frac{18}{12},\,
      2,\,
      3,\,
      5,\,
      10,\,
      \infty
      \right\}.
    \label{eq:probturbdiv.alphavals}
  \end{equation}
\item
  All of our comparisons here begin with this sequence of the $\alpha$ parameter.
\end{textblock}

\begin{textblock}
\item
  In the same fashion as our
  rank-turbulence divergence allotaxonographs~\cite{dodds2023a}---but with some necessary and key modifications---our allotaxonographs
  for probability-turbulence divergence
  pair two complementary visualizations:
\item 
  A map-like histogram and a ranked list.
\end{textblock}

\begin{textblock}
\item
  In isolation, both the histogram and the ranked list have important but limited descriptive power.
\item
  The histogram helps us see how well our choice of $\alpha$ performs,
  information that is entirely lost by the ranking process.
\item
  And the ranked list would be difficult to intuit from the histogram alone.
\end{textblock}

\begin{textblock}
\item
  Many aspects of our allotaxonographs are configurable.
  On
  \href{\gitlaburl}{GitLab},
\item
  we provide our universal code for generating allotaxonographs for
  rank-turbulence divergence, probability-turbulence divergence,
\item
  and other probability divergences
  (see Sec.~\ref{sec:probturbdiv.methods.code}).
\end{textblock}

\begin{textblock}
\item
  We complement all of our allotaxonographs with PDF flipbooks
  which move systematically through a range of $\alpha$ values.
\item
  These flipbooks can be found in the paper's
  supplementary material on
  \href{\zenodourl}{Zenodo}~\cite{dodds_PTD_flipbooks_2025},
\item 
  the paper's \onlineappendices,
\item 
  and
  the associated GitLab repository
  (\gitlablink).
\end{textblock}

\begin{table*}[t!]

  \centering
  
  \renewcommand{\arraystretch}{1.3}

  \small

  \rowcolors{2}{gray!5}{gray!15}  
  \begin{tabular}{l}

    \hline

    \rowcolor{gray!25}
    \textbf{Sites for supplementary material:}

    \\
    \hline

    Allotaxonometry home:

    \onlineappendiceslink

    \\[2pt]

    GitLab repository:

    \gitlablink
    
    \\[2pt]
    
    Zenodo record:

    \zenodolink~\cite{dodds_PTD_flipbooks_2025}

    \\[5pt]

    \hline

    \rowcolor{gray!25}
    \textbf{Direct links to flipbooks on Zenodo:}

    \\
    \hline
    
    Flipbook S1---Pride and Prejudice, first half versus second half, 1-grams:

    \\
    
    \href{
      \zenodofileslink/allotaxonometer-flipbook-1-probability-divergence-pride-and-prejudice-1-grams.pdf?download=1%%
    }{
      \zenodofileslink/allotaxonometer-flipbook-1-probability-divergence-pride-and-prejudice-1-grams.pdf
    }

    \\

    Flipbook S2---Pride and Prejudice, first half versus second half, 2-grams:

    \\

    \href{
      \zenodofileslink/allotaxonometer-flipbook-2-probability-divergence-pride-and-prejudice-2-grams.pdf?download=1%%
    }{
      \zenodofileslink/allotaxonometer-flipbook-2-probability-divergence-pride-and-prejudice-2-grams.pdf
    }

    \\

    Flipbook S3---Pride and Prejudice, first half versus second half, 3-grams:

    \\

    \href{
      \zenodofileslink/allotaxonometer-flipbook-3-probability-divergence-pride-and-prejudice-3-grams.pdf?download=1%%
    }{
      \zenodofileslink/allotaxonometer-flipbook-3-probability-divergence-pride-and-prejudice-3-grams.pdf
    }

    \\

    Flipbook S4---Twitter, 2020-03-12 versus 2020-05-30, 1-grams:

    \\

    \href{
      \zenodofileslink/allotaxonometer-flipbook-4-probability-divergence-twitter-1-grams.pdf?download=1%%
    }{
      \zenodofileslink/allotaxonometer-flipbook-4-probability-divergence-twitter-1-grams.pdf
    }

    \\

    Flipbook S5---Twitter, 2020-03-12 versus 2020-05-30, 2-grams:

    \\

    \href{
      \zenodofileslink/allotaxonometer-flipbook-5-probability-divergence-twitter-2-grams.pdf?download=1%%
    }{
      \zenodofileslink/allotaxonometer-flipbook-5-probability-divergence-twitter-2-grams.pdf
    }

    \\

    Flipbook S6---Twitter, 2020-03-12 versus 2020-05-30, 3-grams:

    \\
    
    \href{
      \zenodofileslink/allotaxonometer-flipbook-6-probability-divergence-twitter-3-grams.pdf?download=1%%
    }{
      \zenodofileslink/allotaxonometer-flipbook-6-probability-divergence-twitter-3-grams.pdf
    }

    \\

    Flipbook S7---Barro Colorado Island, 1985 versus 2015 census, species counts:

    \\

    \href{
      \zenodofileslink/allotaxonometer-flipbook-7-probability-turbulence-divergence-barro-colorado-island.pdf?download=1%%
    }{
      \zenodofileslink/allotaxonometer-flipbook-7-probability-turbulence-divergence-barro-colorado-island.pdf
    }

    \\
    \hline

  \end{tabular}
  
  \caption{
    List of supporting material sites and direct links for probability-turbulence divergence.
    Flipbooks show how probability-turbulence divergence behaves
    as $\alpha$ ranges from 0 to $\infty$
    for the three studies in this paper:
    Pride and Prejudice,
    Twitter,
    and
    Barro Colorado Island.
  }
  \label{tab:probturbdiv.links}
\end{table*}

\subsection{Allotaxonographs for Pride and Prejudice}
\label{sec:probturbdiv.janeausten}

\begin{textblock}
\item
  For a primary, familiar example to help us explain our
  probability-turbulence divergence allotaxonographs,
  we compare the normalized usage frequency distributions of $2$-grams
  between
  the first and second halves of Pride and Prejudice~\cite{austen2001a}.
\item
  We emphasize that we are mostly intent here on showing how allotaxonographs function.
\item
  We are not attempting to reveal astonishing insights into one of
  the most well-regarded
  and well-studied novels of all time.
\end{textblock}

\begin{textblock}
\item
  In Flipbooks
  \href{\zenodofileslink/allotaxonometer-flipbook-1-probability-divergence-pride-and-prejudice-1-grams.pdf?download=1}{S1},
  \href{\zenodofileslink/allotaxonometer-flipbook-2-probability-divergence-pride-and-prejudice-2-grams.pdf?download=1}{S2},
  and
  \href{\zenodofileslink/allotaxonometer-flipbook-3-probability-divergence-pride-and-prejudice-3-grams.pdf?download=1}{S3}
  in the paper's supplementary material~\cite{dodds_PTD_flipbooks_2025},
\item
  we give the reader a view into
  how  our allotaxonometric comparisons
  of the usage frequencies
  of 1-grams, 2-grams, and 3-grams
  in the two halves of Pride and Prejudice
  behave as a function of $\alpha$
\item 
  (see Sec.~\ref{sec:probturbdiv.methods.suppmaterial}).
\item
  All of the flipbooks
  are especially informative
  in showing how contour lines and
  ranked lists change with $\alpha$.
\item
  Here in the main paper,
  we present and discuss five $\alpha$ values,
  all of which tell a story.
\end{textblock}

\begin{revisionbar}

\subsubsection{Focus on most rare types: $\alpha$=0}
\label{subsubsec:probturbdiv.janeausten-0}

\begin{textblock}
\item
  To help demonstrate how tuning $\alpha$ affects the ranked list
  of dominant contributions,
\item 
  in Fig.~\ref{fig:probturbdiv.allotaxonometer9000-jane-austen-2-grams-0}
\item 
  we start with an allotaxonograph for
  $\alpha = 0$
  for the first and second halves of Pride and Prejudice.
\item
  Much of the present section describes allotaxonographs for PTD in general.
\end{textblock}

\begin{textblock}
\item
  First, we describe the format of our allotaxonographs.
\item
  The histogram on the left bins all pairs
\item 
  $
  (
  \logten
  p_{\elementsymbol,1}
  ,
  \logten
  p_{\elementsymbol,2}
  ).
  $
\item
  The shape we see here is typical of comparisons between systems with heavy-tailed
  size-rank distributions.
\item 
  We rotate the axes so as not to privilege one system over the other,
\item 
  as this might lead to a false
  sense of an independent-dependent variable relationship~\cite{bergstrom2018a,dodds2023a}.
\end{textblock}

\begin{textblock}
\item
  We indicate counts per cell using the perceptually uniform colormap magma~\cite{liu2018a}.
\item 
  Because of the logarithmic scale, the cells start to separate for lower values of probability,
  corresponding to counts of 0, 1, 2, and so on.
\item 
  We emphasize that for each cell, the counts as indicated by color
  refer to the number of types (not tokens) that
  have a pair of probabilities in the two systems that fall
  within that cell.
\item 
  For example, the 2-gram ``of the'' is extremely common in both the first and
  second halves of Pride and Prejudice, and is slightly more common
  in the second half. 
\item
  There are no other 2-grams that have a pair
  of probabilities that are the same or close enough to ``of the'''s pair,
\item 
  and hence it is represented with the lightest color indicating a count of 1.
\end{textblock}

\begin{textblock}
\item 
  Types that are common in both systems will be located toward the top of the histogram,
  those that are rare in both will be at the bottom,
\item 
  and those appearing more prevalently in one system will appear further
  away from the vertical midline.
\item 
  Exactly how much this latter category matters is a function of the divergence at hand.
\end{textblock}

\begin{textblock}
\item
  The most extreme cells can be readily understood.
\item
  The bottommost pair of cells represent all 2-grams that appear once in one half of Pride and Prejudice
  and zero times in the other---the novel's 2-gram \textit{hapax legomena}.
\item 
  The lowest cell along the centerline contains 2-grams that appear exactly once in
  each half of Pride and Prejudice.
\end{textblock}

\begin{textblock}
\item
  We annotate example 2-grams around the edges of the histogram.
\item 
  These annotations are more than just decoration---for all divergences we are familiar with,
  types at the edges of the histogram are the only
  ones that can dominate the overall divergence measure.
\item 
  The cells along the upper curved edges largely contain only one 2-gram each (pale yellow).
\item 
  Along the bottom edges, parallel to the axes, edge cells may contain many 2-grams,
\item 
  and the examples shown are random selections.
\end{textblock}

\begin{textblock}
\item
  For any setting of $\alpha$, the contour lines on the histogram
  will indicate
  where $\probdivelement{\alpha}$ is constant.
\item
  For $\alpha$=$0$---which is a special case for the contour lines---we see that the lines do not match well
  with the edges of the histogram.
\item 
  We also see how $\alpha$=$0$ gives weight only
  to system-exclusive 2-grams
\item 
  (i.e.,
  2-grams that appear in only one of the two halves).
\end{textblock}

\begin{textblock}
\item
  As we showed in Sec.~\ref{subsec:probturbdiv.probturbdivzero},
\item 
  when $\alpha$=0,
  the divergence contribution for all types that appear in
  both systems is
  $\probdivelement{0}=0$,
\item 
  while for all exclusive types,
\item 
  $
  \probdivelement{0}
  =
  \frac{1}{
    \left(
    \Ntypesa + \Ntypesb
    \right)
  }
  $
\item 
  (as reflected in the equal bars in the ranked list).
\item 
  Thus, the vertical contour lines
  in Fig.~\ref{fig:probturbdiv.allotaxonometer9000-jane-austen-2-grams-0},
\item 
  which again are present because we anchor them at evenly spaced locations
  along the bottom of the main histogram,
\item 
  all correspond to a divergence of
  $\probdivelement{0}$=$0$.
\item 
  The dashed parts of the visible contour lines
  then collapse to the bottom zero point,
\item 
  showing how the exclusive types
  provide the only non-zero contributions to
  $\probdivelement{0}$.
\end{textblock}

\begin{textblock}
\item
  For all of our allotaxonographs,
  we place 10 contour lines on each half of the histogram's diamond.
\item
  These contour lines are evenly spaced not by height but
\item 
  are rather anchored to the bottom axes of the main histogram
  where they are evenly spaced in logarithmic space.
\end{textblock}

\begin{textblock}
\item
  While it may appear that we have omitted annotations internal to the histogram
  for convenient purposes
  of visualizing the histogram more cleanly, our annotations are intentional.
\item 
  Because individual 2-grams internal to the histogram will never dominate standard divergences,
  highlighting them would be badly misleading~\cite{monroe2008a,kessler2017a}.
\item 
  For our allotaxonographs, the annotations along the bottom of the histogram potentially
  fall into this trap: 
\item
  `rejecting him' and `escape all',
\item
  each appearing once overall,
\item 
  are just two examples of tens of thousands of 2-gram \textit{hapax legomena}.
\item 
  Such types may matter in aggregate but not individually.
\end{textblock}

\begin{textblock}
\item
  Now, our allotaxonographs for probability-turbulence divergence
  must depart from those of rank-turbulence divergence because we
  have to accommodate instances of $\logten p$ when $p = 0$.
\item 
  For ranks, types with 0 counts in one system---exclusive types---are
  assigned a tied rank for last place, necessarily a finite number.
\item 
  Here, on a logarithmic scale our exclusive types would have to be 
  located on one axis at $\logten p = -\infty$.
\item 
  We end the main histogram's domain
  for the lower value of
\item 
  $\logten p_{\elementsymbol,i}$
  such that
  $p_{\elementsymbol,i} > 0$.
\item 
  We then add lighter-colored regions to the bottom of both sides of
  the histogram, and locate
\item 
  $p_{\elementsymbol,i} = 0$
\item 
  along their midlines.
\item 
  The transition is a discrete jump (we do not smoothly interpolate),
  and we connect the contour lines with a dotted line.
\end{textblock}

\begin{textblock}
\item
  The last piece for the histogram is the list of balances
  at the bottom right.
\item 
  These summary quantities are intended to
  be both informative and diagnostic,
\item 
  and they have important if subtle differences.
\item 
  The first bars show the balance of total 2-gram counts
  which for our example of Pride and Prejudice is 50/50 by construction.
\item 
  The second and third balances refer to sizes of lexicons,
  and these are well balanced too.
\item 
  If we create a lexicon of all 2-grams for Pride and Prejudice,
  about 58.2\% of them appear in the first half and 58.3\% in the second.
\item 
  If we instead create separate lexicons of 2-grams for the two halves
  of Pride and Prejudice,
\item 
  the third line of the balances records
  the percentage of 2-grams that are exclusive to each half.
\end{textblock}

\begin{textblock}
\item
  While it could be said that we are ultimately creating a simple 2-d histogram 
  for a joint probability distribution with heavy tails,
\item 
  to our knowledge, there have been relatively few other attempts to do so~\cite{monroe2008a,kessler2017a}.
\item 
  As we have described them, we believe our histograms are crafted
  with a number of special details that make them well suited to their task.
\end{textblock}

\begin{textblock}
\item
  The ranked list on the right maps the two dimensions of the histogram
\item
  onto an ordered single dimension of divergence contributions, largest first.
  The left-right arrangement is solely done to be consistent
  with the histogram---all contributions are positive.
\item 
  The light gray numbers opposite each 2-gram
  (e.g., \textcolor{grey}{$44,185 \rightleftharpoons 433.5$} for
  `the Parsonage')
  indicate the
  2-gram's rank in the first and second halves
  of Pride and Prejudice
\item 
  (and in general, $\systema$ and $\systemb$).
\item 
  We note that because we use tied ranks,
  types will have ranks that are either integers or half-integers~\cite{dodds2023a}.
\end{textblock}

\begin{textblock}
\item
  In the ranked list, we add open triangles to types if they
  are exclusive to one system,
  corresponding to those appearing in the zero probability expansions
  of the histogram.
\item
  For example, `to Brighton' appears only in the first half
  of Pride and Prejudice, and `the Parsonage' only in the second.
\end{textblock}

\begin{textblock}
\item
  In reducing such a high-dimensional categorical space---where
  each unique type represents a dimension---we have first
  collapsed the data to a 2-d histogram, and then to a 1-d list.
\item 
  Being able to find the shape in the histogram to which we can
  apply an instance of probability-turbulence divergence
  gives us some suggestive proof in the pudding.
\end{textblock}

\subsubsection{Squared-chord distances: $\alpha$=1/2}
\label{subsubsec:probturbdiv.janeausten-1/2}

\begin{textblock}
\item
  Fig.~\ref{fig:probturbdiv.allotaxonometer9000-jane-austen-2-grams-1/2}
  shows how PTD changes as we jump from
  $\alpha$=0
  to
  $\alpha$=1/2.
\item
  The contour lines are now showing some alignment with the histogram.
\item
  We now see 2-grams that appear in both halves in the top contributors list.
\item
  For example,
  `Miss Bingley' and `Sir William'.
\item
  That said, many of the top contributors remain exclusive types, from either half.
\item
  The contour lines trim into the main histogram as they come down,
  and then still show some sharp inward turn for the exclusive types.
\end{textblock}

\begin{textblock}
\item
  We point out that any distance measure of the squared-chord family 
  (Sec.~\ref{subsubsec:probturbdiv.alpha=1/2})
  could be represented by a similar allotaxonograph.
\item
  The contour lines and type ordering would remain the same.
\item
  And the user would be given immediate visual feedback on
  the appropriateness of using such a measure for whatever
  probability distributions they are comparing.
\end{textblock}

\subsubsection{Scale-equalizing fit with $\alpha$=3/4}
\label{subsubsec:probturbdiv.janeausten-3/4}

\begin{textblock}
\item
  In Fig.~\ref{fig:probturbdiv.allotaxonometer9000-jane-austen-2-grams-3/4},
  we show an allotaxonograph with $\alpha$=3/4 which we deem to be scale-equalizing.
\item
  We see that the choice of $\alpha$=3/4 generates a list
  with 2-grams from across the rare-to-common spectrum.
\item 
  The balanced darker shadings of annotations in the histogram
  add further support.
\item
  Of course, $\alpha$=3/4 is in itself rough---we are never going
  to be looking for many significant figures for the scale-equalizing value,
  but rather a small range.
\end{textblock}

\subsubsection{$L_{q}$ distances: $\alpha$=1}
\label{subsubsec:probturbdiv.janeausten-1}

\begin{textblock}
\item
  In Fig.~\ref{fig:probturbdiv.allotaxonometer9000-jane-austen-2-grams-1},
  we step up to the $\alpha$=1 case.
\item
  In comparison to $\alpha$=3/4, we now start to see common 2-grams
  rise to the top of the contribution list.
\item 
  The 2-grams `in the', `had been', and `to be' are the top 3,
  and
  `Miss Bingley' has dropped to 5th.
\item
  In moving from $\alpha$=1/2 to $\alpha$=1, we have
  crossed a threshold of scale-equalizing.
\item
  Again, $\alpha$=3/4 is rough but for the user's interpretation
  of how all scales contribute,
  it provides a sufficiently balanced allotaxonograph.
\end{textblock}

\subsubsection{Focus on most common types: $\alpha$=$\infty$}
\label{subsubsec:probturbdiv.janeausten-infty}

\begin{textblock}
\item
  Finally, in
  Fig.~\ref{fig:probturbdiv.allotaxonometer9000-jane-austen-2-grams-infty},
  we show the form of the allotaxonograph for the extreme of $\alpha$=$\infty$.
\item
  The most common types now dominate
  (providing they do not have the
  same probability in each system, itself a very unlikely event for real systems).
\item
  The contour lines are now parallel to the upper borders of the diamond.
\end{textblock}

\end{revisionbar}

\subsection{Allotaxonographs for Twitter}
\label{sec:probturbdiv.twitter}

\begin{textblock}
\item
  For our second set of allotaxonographs,
  we compare two key dates of two major events through the lens of English-speaking Twitter:
\item 
  2020/03/12, the date that COVID-19 became the major story in the United States,
\item 
  and
  2020/05/30,
\item 
  five days after the murder of George Floyd in Minneapolis, Minnesota,
  by
  police officer Derek Chauvin~\cite{wu2021a}.
\item 
  We compare day-scale normalized usage frequency
  distributions for 1-, 2-, and 3-grams for these two dates in 
\item 
  Figs.~\ref{fig:probturbdiv.allotaxonometer9000-2020-03-12-2020-05-30-story-wrangler-1grams-all-prob-div},
  \ref{fig:probturbdiv.allotaxonometer9000-2020-03-12-2020-05-30-story-wrangler-2grams-all-prob-div},
  and
  \ref{fig:probturbdiv.allotaxonometer9000-2020-03-12-2020-05-30-story-wrangler-3grams-all-prob-div}.
\item
  These datasets are far larger than Pride and Prejudice
  with approximately $10^{7.5}$ types
  and $10^{8.5}$ tokens.
\item 
  Nevertheless, producing an allotaxonograph on a standard present-day laptop
  takes on the order of minutes,
\item 
  which is almost entirely accounted
  for by loading and merging of the distributions.
\end{textblock}

\begin{textblock}
\item
  We choose 2020/03/12 as a key date for the COVID-19 pandemic for
  several reasons. 
\item
  First and primarily, the World Health Organization
  (WHO) officially declared the COVID-19 outbreak to be a pandemic on
  2020/03/11, 
\item
  a decision that was amplified immediately online but
  discussion of which most strongly appeared in the news and on Twitter
  on the following day.
\end{textblock}

\begin{textblock}
\item
  The date of 2020/03/11 also saw a confluence of three major events that jolted the United States
  and dramatically elevated the story of the pandemic,
\item 
  all occurring tightly around a 15-minute period between 9 and 10 pm EDT (1 am to 2 am UTC).
\item 
  First, the National Basketball Association (NBA)
  abruptly suspended its season.
\item 
  The central event was the abandoning of a game just before tipoff between the Utah Jazz and Oklahoma City Thunder,
\item 
  upon the league learning that Rudy Gobert, a center for the Utah Jazz, had tested positive for COVID-19.
\item 
  Other players would test positive in the coming days and weeks, as would staff for teams and members
  of the media.
\item 
  Just a few days earlier, Gobert had joked with the media about his perception of
  institutional overreaction to the coronavirus by touching microphones at an interview.
\end{textblock}

\begin{textblock}
\item
  Second, Tom Hanks announced that both he and his wife Rita Wilson had
  tested positive for COVID-19 while Hanks was working on a Baz Luhrmann film in Australia.
\item 
  Hanks was at the time the most high-profile figure known to have contracted COVID-19.
\end{textblock}


\begin{textblock}
\item
  Third, President Donald Trump gave an Oval Office Address, the second of his presidency,
  ``On the Coronavirus Pandemic.''
\item 
  The address marked a strong shift in Trump's rhetoric regarding the danger of the COVID-19 outbreak.
\item 
  The main decision announced was the ban on travel from Europe to the US for 30 days,
\item 
  which was later clarified to not also mean a ban on trade.
\item 
  Futures on the US stock market dropped during the speech.
\end{textblock}

\begin{textblock}
\item
  Combined, these disparate events were a major part of
  the COVID-19 pandemic becoming the dominant story for what would become weeks, months, and then years ahead.
\end{textblock}

\begin{textblock}
\item
  The murder of George Floyd on 2020/05/25, Memorial Day in the US, 
\item
  precipitated Black Lives Matter protests and civilian-police confrontations in Minneapolis.
\item 
  The protests would grow over the following weeks, and begin to spread around the world.
\item 
  And, at least in the first week, George Floyd's murder overtook coronavirus as the dominant story
  in the US~\cite{dodds2021d}.
\end{textblock}

\begin{textblock}
\item 
  With the above context in mind,
\item
  we can sensibly examine the allotaxonographs of
\item
  Figs.~\ref{fig:probturbdiv.allotaxonometer9000-2020-03-12-2020-05-30-story-wrangler-1grams-all-prob-div},
\item
  \ref{fig:probturbdiv.allotaxonometer9000-2020-03-12-2020-05-30-story-wrangler-2grams-all-prob-div},
\item
  and
  \ref{fig:probturbdiv.allotaxonometer9000-2020-03-12-2020-05-30-story-wrangler-3grams-all-prob-div}.
\end{textblock}

\begin{textblock}
\item
  Our primary observation is that the three histograms vary considerably
  as we move through 1-, 2-, and 3-grams.
\item 
  The histograms broaden with increasing $n$, with the 3-gram
  histogram losing a scaling form and squaring up in the axes.
\item 
  The rapidly growing combinatorial possibilities of $n$-grams
  with increasing $n$
\item 
  mean that we see more and more exclusive
  $n$-grams as we look across the three allotaxonographs.
\item 
  For 1-grams, around 60\% of each date's lexicon are exclusive,
\item 
  for 2-grams, the percentage increases to around 70\%,
\item 
  and
\item 
  for 3-grams we reach 80\% 
\item
  (see the bottom of the three
  balance summaries in each allotaxonograph).
\end{textblock}

\begin{textblock}
\item
  The maximum count per cell
  is $10^{6}$ for 1-grams,
\item 
  $10^{7}$ for 2-grams,
\item 
  and
  $10^{8}$ for 3-grams.
\item 
  The cells with the most $n$-grams are of course
  the \textit{hapax legomena}---the bottommost two cells in
  the histogram---those $n$-grams which
  appear once on one of the dates and not at all on the other.
\end{textblock}

\begin{textblock}
\item
  To obtain good balance for the most dominant $n$-grams,
  we select $\alpha$=1/3, 5/6, and $\infty$.
\item 
  Different kinds of terms dominate depending on $n$
  with
\item 
  `coronavirus',
  `the coronavirus',
  and
  `tested positive for'
  leading on 2020/03/12,
\item 
  and
  `Minneapolis',
  `George Floyd',
  and
  `of George Floyd'
  at the top on 2020/05/30.
\end{textblock}

\begin{textblock}
\item
  Because social amplification is encoded
  in Twitter's data stream through retweets,
\item 
  dominant 2-grams and especially 3-grams are liable
  to belong to the most retweeted messages of the day,
  and may lead to some variation in the dominant $n$-grams.
\item 
  (By contrast, we do not have a measure of
  popularity of individual phrases or sentences
  within Pride and Prejudice with just the bare text.)
\item 
  For example, `toilet paper'
  and
  `World Health Organization'
  appear as dominant 2-grams and 3-grams
\item
  but none of their five
  distinct 1-grams are near the top of the ranked list
  in
  Fig.~\ref{fig:probturbdiv.allotaxonometer9000-2020-03-12-2020-05-30-story-wrangler-1grams-all-prob-div}.
\item 
  On the other hand, some dominant 1-grams may be used in diverse 2-grams and 3-grams
  and thus may not 
  appear in the ranked lists for 2-grams and 3-grams.
\item 
  Examples from
  Fig.~\ref{fig:probturbdiv.allotaxonometer9000-2020-03-12-2020-05-30-story-wrangler-1grams-all-prob-div}
  are
  `antifa' and `Breonna'.
\end{textblock}

\begin{textblock}
\item
  For all three $n$-gram comparisons of these two dates on Twitter,
  we provide Flipbooks
\item 
  \href{\zenodofileslink/allotaxonometer-flipbook-4-probability-divergence-twitter-1-grams.pdf?download=1}{S4},
  \href{\zenodofileslink/allotaxonometer-flipbook-5-probability-divergence-twitter-2-grams.pdf?download=1}{S5},
  and
  \href{\zenodofileslink/allotaxonometer-flipbook-6-probability-divergence-twitter-3-grams.pdf?download=1}{S6}~\cite{dodds_PTD_flipbooks_2025}
  (Sec.~\ref{sec:probturbdiv.methods.suppmaterial}).
\item
  Readers may use these to easily explore how the choice of $\alpha$ affects
  the fit for the contour lines in the histogram
  and the ordering of which $n$-grams dominate
  probability-turbulence divergence.
\end{textblock}

\subsection{Allotaxonographs for Barro Colorado Island}
\label{sec:probturbdiv.BCI}

\begin{textblock}
\item
  We include one final allotaxonograph from an entirely different field of research,
  ecology.
\item 
  In Fig.~\ref{fig:probturbdiv.allotaxonometer9000-1985-2015-barro-colorado002}
  we show a probability-turbulence divergence allotaxonograph for
  tree species abundance on Barro Colorado Island
  for censuses
  completed in 1985 and 2015.
\item 
  This example also shows how allotaxonographs can be used to
  inspect how well divergence measures perform for datasets
  that are much smaller than our examples from literature and Twitter.
\item 
  The species that dominates the overall divergence score
  is one that has diminished in abundance,
\item 
  \textit{Piper cordulatum}~\cite{trelease1927a,standley1927a,thies2004a,andrade2013a}.
\item 
  In Ref.~\cite{dodds2023a}, we compared these distributions with
  rank-turbulence divergence,
\item 
  and the overall orderings of dominant species
  are broadly consistent.
\end{textblock}

\begin{textblock}
\item
  In Flipbook
  \href{\zenodofileslink/allotaxonometer-flipbook-7-probability-turbulence-divergence-barro-colorado-island.pdf?download=1}{S7},
\item 
  we show how the dominant contributions of species
  vary as a function of $\alpha$.
\end{textblock}

\section{Data, Code, and Supplementary Material}
\label{sec:probturbdiv.methods}

\subsection{Datasets}
\label{sec:probturbdiv.methods.datasets}

\begin{textblock}
\item
  \textbf{Pride and Prejudice:}
\item 
  We sourced a plain text version of Jane Austen's
  Pride and Prejudice from Project Gutenberg
\item 
  (\href{http://www.gutenberg.org/ebooks/1342}{http://www.gutenberg.org/ebooks/1342}).
\end{textblock}

\begin{textblock}
\item
  \textbf{Normalized $n$-gram usage frequency on Twitter:}
\item 
  We collected around 10\% of all tweets sent on these dates based on
  Coordinated Universal Time (UTC), meaning they covered
  8:00:00 pm to 7:59:59 pm Eastern Daylight Time (EDT)
  and 5:00:00 pm to 4:59:59 pm Pacific Daylight Time (PDT).
\item 
  We provide historical access to the top $10^{6}$
  1-grams, 2-grams, and 3-grams across more than 100 languages
  as part of our
  Storywrangler for Twitter project~\cite{alshaabi2021c}.
\end{textblock}

\begin{textblock}
\item
  \textbf{Species abundance on Barro Colorado Island:}
\item 
  We accessed the dataset for BCI censuses performed roughly every 5 years over 35 years
  through the online repository described in Ref.~\cite{condit2019a}.
\end{textblock}

\subsection{Code for divergence calculation and rendering allotaxonographs}
\label{sec:probturbdiv.methods.code}

\begin{textblock}
\item
  All scripts and documentation reside on
  \href{https://gitlab.com/}{GitLab}
  at
  \gitlablink.
\item
  For the present paper, we wrote the scripts to generate
  the allotaxonographs in MATLAB.
\item 
  We originally produced all figures and flipbooks using MATLAB Version R2020a,
  while endeavoring to keep the code functional with future versions.
\item 
  As is, the script also generates allotaxonographs
  for rank-turbulence divergence~\cite{dodds2023a},
  the present paper's probability-turbulence divergence,
  and a parametrized symmetric entropy divergence
  that generalizes Jensen-Shannon divergence.
\item 
  In Ref.~\cite{st-onge2025a},
  we report our contributions for
  Python and browser-based allotaxonographs
  for rank-turbulence divergence.
\item 
  We welcome ports of our general allotaxonometry framework
  to other languages~\cite{st-onge2025a}.
\end{textblock}

\subsection{Supplementary material}
\label{sec:probturbdiv.methods.suppmaterial}

\begin{textblock}
\item
  We list all relevant sites and links in Tab.~\ref{tab:probturbdiv.links}.
\item
  The base site \onlineappendiceslink\ provides a home for our work on allotaxonometry.
\item
  We use \zenodobaseurl\ to store flipbooks~\cite{dodds_PTD_flipbooks_2025},
  and we give
  direct links to all flipbooks in Tab.~\ref{tab:probturbdiv.links}.
\item
  Flipbooks are designed to be examined with a PDF viewer in single-page mode.
\item
  Some PDF viewers within browsers do not accommodate single-page mode,
  and we recommend using an alternate browser
  or downloading flipbooks and viewing locally.
\item
  The same flipbooks are available on the base site and in our
  \href{\gitlaburl}{GitLab repository}.
\end{textblock}

\section{Concluding remarks}
\label{sec:probturbdiv.concludingremarks}



\begin{textblock}
\item
  \revision{
    We have defined, analyzed, and demonstrated the use
    of probability-turbulence divergence as an instrument
    of allotaxonometry,
    both analytically and through the diagnostic visualizations
    afforded by our allotaxonographs.
  }
\item 
  As the probability-based analog of
  our rank-turbulence divergence,
\item 
  the instrument is able to perform well
  when comparing heavy-tailed size-rank distributions
  of type frequencies.
\item
  \revision{
    We emphasize the importance
    of creating allotaxonographs for a range of
    values of $\alpha$ (see Eq.~\ref{eq:probturbdiv.alphavals}).
    }
\item
  From there, the user can assess whether or not
  a scale-equalizing value of $\alpha$ exists.
\item 
  We have shown further that probability-turbulence divergence
  generalizes a range of existing probability-based divergences,
\item 
  either matching in exact form or equating in how types
  are ordered by type contribution.
\end{textblock}

\begin{textblock}
\item
  While we view rank-turbulence divergence as our most general, interpretable instrument,
\item 
  for systems in which probabilities (or rates) of types occurring are
  well defined---and the resulting distributions involved are
  heavy-tailed---probability-turbulence divergence provides a more
  nuanced instrument.
\end{textblock}

\begin{textblock}
\item
  We also favor divergences which compare distributions in
  as transparent a way as possible.
\item 
  To that end, we have made the core of probability-turbulence divergence
  a simple difference of powers of probabilities
  (Eq.~\ref{eq:probturbdiv.probdiv_definition_core}).
\item 
  By contrast, we view some divergences as being problematic
  in being overly constructed.
\item 
  We venture that Jensen-Shannon divergence (JSD), which we ourselves have used elsewhere,
  is one such instrument.
\item 
  The creation of an artificial mixed distribution
  is a contrivance we avoid here,
\item 
  and is perhaps indicative of taking information theory too far~\cite{shannon1956a}.
\end{textblock}

\begin{textblock}
\item
  In our experience, we have also found that the visual information
  delivered by our allotaxonographs,
\item 
  especially in their coupling of histograms and ranked lists,
\item 
  has been essential to working effectively with divergences of all kinds.
\end{textblock}

\begin{textblock}
\item
  One caution we make is that in the examples we have explored in the present paper,
  we have taken distributions as they are.
\item
  That is, we have not contended with issues of sub-sampling
  and missing tail data~\cite{haegeman2013a}.
\item
  We can say that types appearing with a high rate (e.g., common $n$-grams on
  Twitter) will not be affected by accessing more data, as they are
  well-estimated rates.
\item
  In our paper on rank-turbulence divergence, we examined
  how truncation of distributions affects allotaxonographs,
  and such an approach is always available for any divergence.
\end{textblock}

\begin{textblock}
\item
  Finally, in our present paper and in Ref.~\cite{dodds2023a},
  we have so far made choices of $\alpha$ based on inspection of the relevant histogram.
\item 
  A clear next step is to find ways to determine a scale-equalizing $\alpha$ for any given pair of distributions,
  and to do so only when sufficiently robust scaling is apparent.
\item 
  From a storytelling perspective, we are concerned with finding
  a scale-equalizing $\alpha$ that returns
  a ranked list of distinguishing types for two distributions such that the list comprises
  a balance of types from across the full range of observed probabilities~\cite{williams2015b}.
\item 
  We have performed some preliminary work for such an optimization, and note here
  that simple regression is made difficult by the overwhelming weight of rare types
  relative to common ones.
\end{textblock}



\section*{Acknowledgments}
The authors were in part supported by
National Science Foundation Award \#2242829,
a gift from the
Massachusetts Mutual Life Insurance Company,
and
an anonymous philanthropic gift.




\clearpage

\section*{Changelog}
\begin{changelogbox}[2024/07/20]
\item 
  Added this changelog.
\item
  Updated authorship.
\end{changelogbox}

\begin{changelogbox}[2024/09/29]
\item 
  Added three small clarifying sections.
\end{changelogbox}

\begin{changelogbox}[2025/03/14]
\item
  Restructured paper in manuscript format.
\item
  Added logline.
\item
  Added key words
\item
  Added boxes for logline, abstract, keywords, and changelog.
\item
  Added table of contents.
\item
  Rebuilt internal \LaTeX\ structure with text blocks.
\item
  Made minor edits for clarity.
\item
  Updated citations.
\item
  Updated affiliations.
\item
  Improved format of~Tab.~\ref{tab:probturbdiv.connections}.
\item 
  Housed the paper's 7 supplementary flipbooks at \zenodolink~\cite{dodds_PTD_flipbooks_2025}.
\item
  Linked flipbooks directly to files on
  \zenodolink.
\item
  Added a clear list of all supplementary material
  with Tab.~\ref{tab:probturbdiv.links}.
\item
  Added notes on supplementary material
  in Sec.~\ref{sec:probturbdiv.methods.suppmaterial}.
\end{changelogbox}

\begin{changelogbox}[2025/08/18]
\item
  Reframed to state that full range of $\alpha$ is important,
  and flipbooks of allotaxonographs are key to exploring
  any comparison of two systems.
\item
  Provided more explanation leading to the definition
  of probability-turbulence divergence.
\item
  Instead of optimal $\alpha$, now have scale-equalizing $\alpha$.
\item
  Adjusted 
  `categorical distribution' $\rightarrow$ `type-probability distribution'
  in title and throughout the paper.
\item
  Clarified use of frequency as normalized frequency.
\item
  Swapped order of sections on allotaxonographs and connections to extant divergences and distances.
\item
  Expanded section on links to other measures, especially for
  $\alpha$=1/2, 1, and $\infty$.
\item
  Added two more allotaxonographs for Pride and Prejudice,
  for
  $\alpha$=1/2, and 1.
\item
  Rearranged sequence of 5 allotaxonographs for Pride and Prejudice
  to better demonstrate how tuning $\alpha$ affects $\probdiv{\alpha}$.
\end{changelogbox}

\begin{changelogbox}[2026/06/19]
\item
  Adjusted allotaxonograph script so that types appear on left of histogram
  when $\alpha=0$.
\item
  Improved spacing in the Barro Colorado figure (main paper and flipbooks).
\item
  Much copy editing throughout.
\end{changelogbox}

\vfill



\addcontentsline{toc}{section}{References}
\bibliography{probability-turbulence-divergence}







\end{document}